# Directed impulsive molecular dynamics reveal energy-routing pathways of bond cleavage in TATB

*Guanchen Dong*[1,2,4,†], *Xiaona Huang*[5,†], *Yuxuan Chen*[1,3], *Linghua Tan*[2], *Guangcheng Yang*[2,4], *Rui Su*[6], *Lin Wang*[1,*], *Pengfei Guan*[1,*]

1. Advanced Interdisciplinary Science Research (AiR) Center, Ningbo Institute of Materials Technology and Engineering, Chinese Academy of Sciences, Ningbo 315201, China.
2. Nanjing University of Science and Technology, Nanjing, 210094, China.
3. Advanced Structural Ceramics Innovation Center, YongJiang Laboratory, Ningbo 315202, China.
4. China Academy of Engineering Physics·Institute of Chemical Materials, Mianyang, 621900, China.
5. School of Physics and Technology, Nanjing Normal University, Nanjing 210023, China.
6. Institute of Advanced Magnetic Materials, College of Materials & Environmental Engineering, Hangzhou Dianzi University, Hangzhou 310018, China.

[†]These authors contributed equally to this work.

**Corresponding Author**

* Lin Wang (wanglin@nimte.ac.cn); Pengfei Guan (pguan@nimte.ac.cn).

**ABSTRACT**

Responses of energetic materials to external stimuli involve more than uniform thermalization. Understanding directional energy redistribution is essential for structural and ignition design. Here, we use TATB as an example to construct a vibrationally faithful machine-learning potential from only 4,630 configurations. To resolve the elusive nonequilibrium energy flow, we develop directed impulsive molecular dynamics, a statistically controlled protocol that repeatedly deposits equal kinetic energy along selected vibrational modes. By amplifying mode-specific responses while tracking the ensuing nonequilibrium evolution, this approach turns energy redistribution into a directly traceable routing process. We find that initial bond cleavage depends strongly on the initially excited mode, rather than being determined by bond-cleavage barriers alone. Specifically, modes that readily trigger reaction efficiently channel energy towards a bond cleavage coordinate, whereas modes that are less effective at inducing break divert energy into soft-like modes or trap it in modes from which energy flow is inefficient.

Despite continuing efforts to develop more controllable energetic materials, accidental initiation remains a persistent risk,[1] and efficient energy release is still difficult to achieve.[2] Understanding how externally deposited energy is transported, localized and converted into chemical reactions offers provide a route to overcoming these limitations.

Upon exposure to external stimuli, such as shock, heating or electrical excitation, energetic materials first undergo energy localization, followed by the initiation of chemical reactions and subsequently develop into sustained thermal runaway.[3] Hotspot theory[4] establishes the basic picture of energetic material response in terms of local temperature rise and the competition between heat generation and dissipation. During the earliest stage, however, the deposited energy has not yet fully thermalized across the available degrees of freedom. Multiphonon up-pumping theory[5] proposes that mechanical stimulation first excites low-frequency lattice phonons, after which energy is transferred through anharmonic coupling along selective pathways towards higher-frequency intramolecular vibrations. Therefore, energy transfer is not entirely random but may instead be governed by specific transport pathways. Previous microscopic studies of energy transfer about phonon-scattering relationships[6] and anharmonic coupling[7,8] can be used to identify modes that are likely to couple within a given crystal configuration. However, coupling relationships derived from near-equilibrium structures may not directly represent energy-routing pathways during dynamic loading, when the local geometry, intermolecular interactions and vibrational modes continuously evolve.

Among the energetic materials, 2,4,6-triamino-1,3,5-trinitrobenzene (TATB) has been regarded as a representative system for investigating energy-routing pathways because of its relatively high intramolecular symmetry.[9] Experimental investigations using vibrational spectroscopy have provided important insights into the structural response of TATB under external stimuli. Pressure-dependent Raman measurements revealed that compression-induced changes first appear in low-frequency lattice vibrational modes, followed by variations in characteristic molecular vibrations associated with the amino and nitro groups. These spectral changes indicate that mechanical energy is initially accommodated through lattice distortion and intermolecular interaction rearrangement, accompanied by subsequent modification of intramolecular vibrational states before chemical decomposition occurs.[10,11] Besides the experimental techniques, molecular dynamics (MD) simulation offers clear advantages for resolving the microscopic mechanism of energy-routing pathways, such as vibrational modes evolution and initial chemical reactions. Previous ReaxFF-based MD studies have revealed the atomic-scale decomposition processes of TATB under thermal and shock loading, identifying that the initial decomposition is mainly associated with the cleavage of N–H and C–$NO_2$ bonds.[12-14] Meanwhile, recent studies on vibrational energy transfer and intramolecular vibrational energy redistribution under equilibrium conditions have demonstrated that vibrational energy transfer in energetic materials proceeds through specific intermediate modes, highlighting the importance of mode-resolved energy transport in understanding initiation mechanisms.[8,15] Nevertheless, a practical method is still lacking for selectively depositing energy into a specified degree of freedom and directly characterizing its effects on subsequent energy redistribution and bond cleavage.

In this work, a non-equilibrium molecular dynamics method with non-empirical machine learning potential is developed to directly investigate the energy-routing pathways of energetic materials. First, we constructed and validated a neural-network interatomic potential (NNAP) capable of faithfully describing the crystal structure, vibrational characteristics and early reactivity of TATB. On this basis, we developed directed impulsive molecular dynamics (DIMD), which repeatedly deposit energy into vibrational modes defined by instantaneous internal coordinates. This approach enables direct comparison of the ease of bond cleavage and selectivity among different initially excited modes. We further tracked changes in different mode-projected energies to quantify energy transfer to bond-stretching coordinate, retention within the initially excited mode , and redistribution into other modes.

## Results

### A vibrationally faithful NNAP for TATB

TATB combines strong intramolecular covalent bonds with an extensive network of intralayer hydrogen bonds and weak interlayer dispersive interactions. An empirical potential with a fixed functional form may accurately describe selected regions of the TATB potential-energy surface, but it is difficult to achieve simultaneous fidelity for vibrational dynamics and bond rearrangement.[16,17] In our previous work, we developed a framework for constructing NNAP for energetic molecular systems,[18] which has proved effective for efficiently developing high-fidelity NNAP that describes both the structural and reactive responses. Here, we adopt a similar framework to construct a vibrationally faithful NNAP for TATB, as shown in Fig. 1a. Rather than indiscriminately expanding the training set, we organized sampling around the key events that the potential was required to describe. Bias-enhanced sampling enabled efficient exploration of the relevant near-equilibrium structural space, while uncertainty screening and structural-diversity selection were used to progressively enrich the dataset.[19,20] The iterative process of NNAP was therefore guided by target physical properties rather than by test-set errors alone.

Specifically, TATB is characterized by layered molecular packing, an extensive intralayer N–H···O hydrogen-bond network and interlayer H···O contacts that progressively shorten and strengthen under high pressures.[21,22] Given our focus on vibrational motion and energy transfer, the NNAP must accurately preserve structural features and environment-dependent vibrational modes. Near-equilibrium molecular and crystal configurations were used to establish the local covalent geometry and molecular packing. We then performed bias-enhanced hyperactive learning from TATB crystal structures optimized at different pressures, thereby sampling geometric environments associated with distinct packing arrangements. As an iterative convergence criterion, crystal structures were independently optimized using density functional theory (DFT) and the NNAP from each iteration over a hydrostatic-pressure range of 0–50 GPa, and their pressure-dependent structural information was compared. To provide broader coverage of the structural space of TATB, we additionally followed the staged sampling workflow developed in our previous work. The sampling protocol comprised molecular, crystal, pyrolysis and shock stages (Fig. S1a). The structural characteristics of TATB at different pressures rapidly approached convergence and fully satisfied the iterative convergence criterion after 15 iterations (Fig. S1b–d), demonstrating that the resulting NNAP reproduces the key structural features of TATB with DFT-level fidelity.

The final NNAP, trained on a dataset consisting of 4,630 configurations, reproduces the reference DFT-level energy and force benchmarks, as well as the principal structural, dynamic and vibrational benchmarks (Fig. 1b-f). The energy mean absolute error (MAE) of test set is 3.25 meV·atom$^{-1}$ and the force MAE is 147.88 meV·A$^{-1}$. Radial distribution functions and unit cell parameters remain close to DFT or experiment and improve on the compared reactive force fields (ReaxFF)[23] and machine learning potential covering multiple typical energetic materials (EMFF).[24] More importantly for the present work, the infrared (IR) spectrum obtained by NNAP retains the major features reported in DFT and experiment of TATB[25] (Fig. 1f), including the N-H vibrational mode at ~3400 $cm^{-1}$ and C-N vibrational mode at ~ 1100 $cm^{-1}$, which indicates more accurate description of TATB vibrational modes comparing with ReaxFF and EMFF potentials. The comparison potentials show broader shifts and loss of detailed peak structure, indicating that they do not guarantee a faithful vibrational landscape. A separate chemical events benchmark was performed to determine whether the NNAP captures the qualitative difference between the two representative initial bond cleavage channels (Fig. S2). Arrhenius analysis of the temperature-dependent decay of intact bonds yielded apparent activation energies of 286.486 kJ·$mol^{-1}$ for N–H cleavage and 431.447 kJ·$mol^{-1}$ for C–$NO_2$ cleavage. These results establish that C–$NO_2$ cleavage is thermally less accessible than N–H cleavage under conventional sampling.[26]

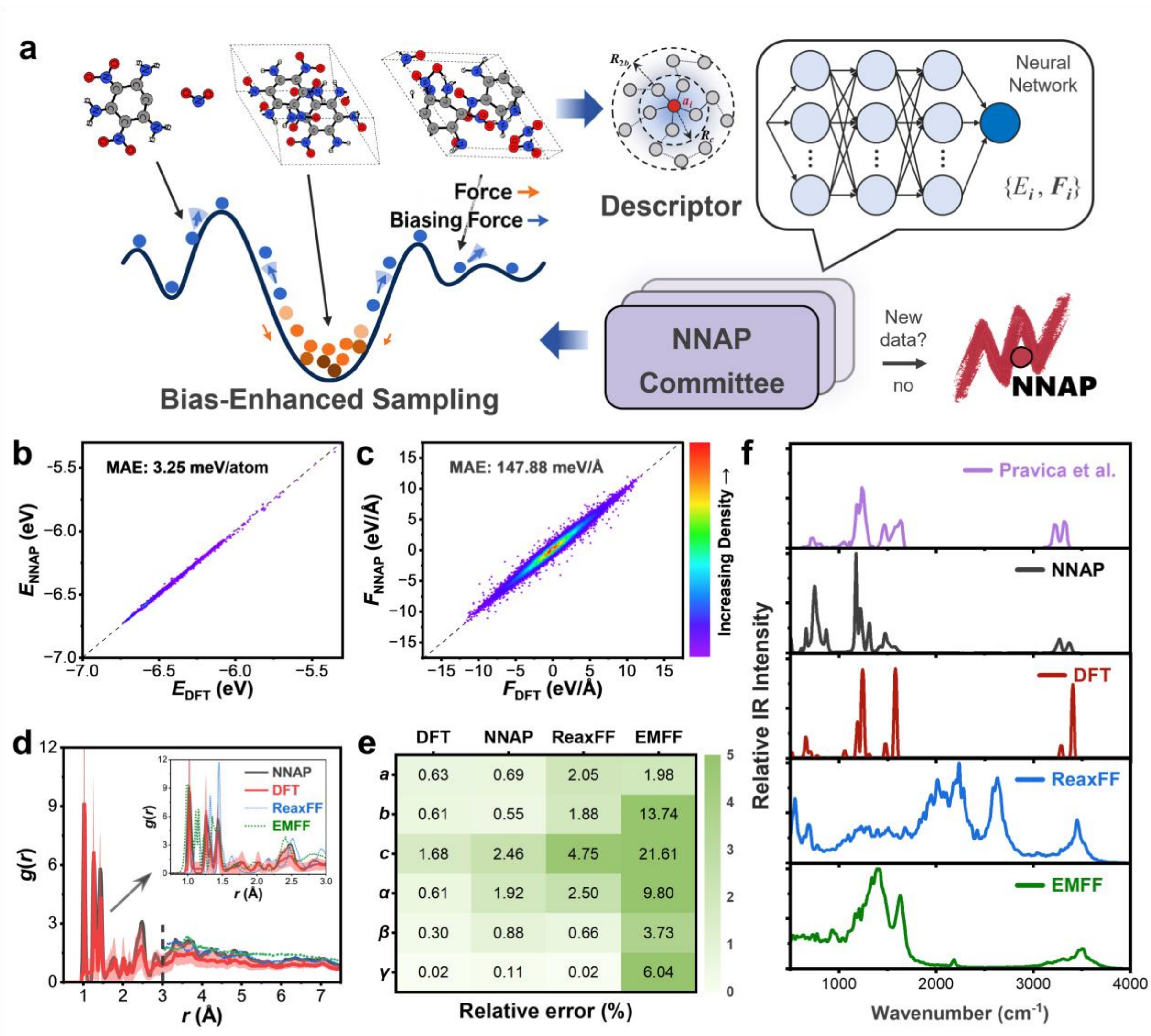

**Fig.** 1 **Bias-enhanced sampling and benchmarks of the NNAP for TATB.** (a) Bias-enhanced sampling NNAP construction strategy combining committee-based screening. The biased MD simulations were based on the committee-averaged NNAP energies and forces. The biased potential energy was defined as the committee-averaged energy reduced by the bias-strength parameter multiplied by the standard deviation of the energies predicted by the individual models in the committee, thereby promoting the exploration of configurations with high model uncertainty.(b, c) Energy and force benchmarks. Comparison of NNAP-predicted energies and atomic forces with the corresponding DFT reference values for an independent test set comprising 10% of the database configurations, which were randomly selected and excluded from model training. (d) RDF calculated using different computational methods from 1 ps NVT trajectories at 300 K with a time step of 0.5 fs. The NNAP reproduces DFT structural correlations

over both short and intermediate ranges. (e) Comparison of the unit-cell parameters with the DFT and experimental values. The NNAP results closely agree with the DFT calculations and deviate from experiment by less than 3%. (f) Comparison of infrared spectra from reference, NNAP and other potentials. The NNAP accurately reproduces the major characteristic peak positions of TATB's IR, including the bands associated with C–N and N–H vibrational modes.

### Controlled mode-selective excitation by directed impulsive molecular dynamics

To clarify the energy-routing pathways after energy enters a particular degree of freedom, we developed DIMD (Fig. 2), drawing on established Hessian-based energy-injection strategies.[27] DIMD represents a prescribed vibrational mode of a probe molecule using a steering vector constructed from its instantaneous internal coordinates, thereby maintaining a physically interpretable connection between the injected energy and the targeted probe molecular motion throughout structural evolution. The steering vector evolves continuously with the molecular geometry, enabling mode-selective energy injection and tracking of subsequent energy redistribution among vibrational modes.

The schematic diagram of DIMD was shown in Fig. 2. An equilibrium MD trajectory is first generated, from which a series of initial atomic coordinates, $\mathbf{r}_{ini}$, and velocities, $\mathbf{v}_{ini}$, are sampled. During the simulation, one TATB was selected as probe molecule. Then eleven vibrational modes that may involve in the energy-routing pathways of initial bond cleavage channels were defined through internal coordinates: rocking, scissoring, torsion and wagging motions of the $NH_2$ and $NO_2$ groups, together with ring-breathing, out-of-plane and trigonal deformations. The schematic diagrams of these modes are provided in Fig. S3. For a prescribed mode $i$, a physically interpretable internal coordinate $\mathbf{Q}_i$ is used to construct its instantaneous Cartesian gradient, $\boldsymbol{g}_i = \left(\frac{\partial \boldsymbol{Q}_i}{\partial r}\right)^T$, where $r$ denotes the 3$N$-dimensional atomic-coordinate vector. The mass-weighted steering vector is defined as $\tilde{\boldsymbol{S}}_i = \boldsymbol{M}^{-1}\boldsymbol{g}_i$, where $\mathbf{M}$ is the diagonal atomic-mass matrix. Normalization gives the instantaneous steering vector $\boldsymbol{S}_i$, which is a prerequisite for the direct comparison of subsequent kinetic energy increments among different selected modes.

After the instantaneous mode vector is selected, energy was injected by increasing the kinetic energy along the instantaneous selected mode. At each impulse, the velocity $\mathbf{v}$ is updated along $\boldsymbol{S}_i$, according to $\mathbf{v}^+ = \mathbf{v}^- + \beta\boldsymbol{S}_i$, where the superscripts - and + denote quantities immediately before and after the impulse, respectively. The instantaneous velocity projection along mode $i$ is $a_i = \boldsymbol{S}_i^T\boldsymbol{M}\boldsymbol{v}$, and $\beta = \mathrm{sign}(a_i^-)\sqrt{(a_i^-)^2 + 2\Delta E} - a_i^-$ is chosen so that the modal kinetic energy increases by exactly $\Delta E$. Between successive energy injections, the system evolves freely, after which $\boldsymbol{S}_i$ is reconstructed from the updated molecular geometry based on $\mathbf{g}_i$.

During the simulation, a single directed excitation is insufficient for resolving the pathways of energy flow, because the deposited energy rapidly disperses into other degrees of freedom within the picosecond timescale. Increasing the impulse energy to enhance the signal could instead induce large nonphysical deformations, while the response of an individual trajectory is strongly influenced by its initial vibrational phase and local configuration. DIMD therefore applies repeated small energy impulses separated by short periods of free evolution, with the steering vector of selected mode updated before each impulse. For each selected mode, 100 independent configurations were sampled from an 800 K equilibrium trajectory of a periodic 3 × 3 × 3 TATB supercell. This temperature was used to accelerate reactive events within the simulation timescale. An energy of $3k_BT$ (0.207 eV at 800 K) was added to the selected mode group every 50 fs. This empirical energy increment was sufficient to generate an observable energy flow without directly inducing bond cleavage. The 50 fs interval was selected based on a velocity-autocorrelation time of approximately 40 fs, allowing energy redistribution while maintaining nonequilibrium evolution. Each impulse was followed by microcanonical dynamics with a 0.1 fs time step. The energy injection was repeated until the first bond cleavage event was detected or 400 segments (20 ps) had elapsed. These Ensemble averaging over independent initial configurations then reveals statistically mode-dependent energy flow tendencies while maintaining controlled nonequilibrium excitation.

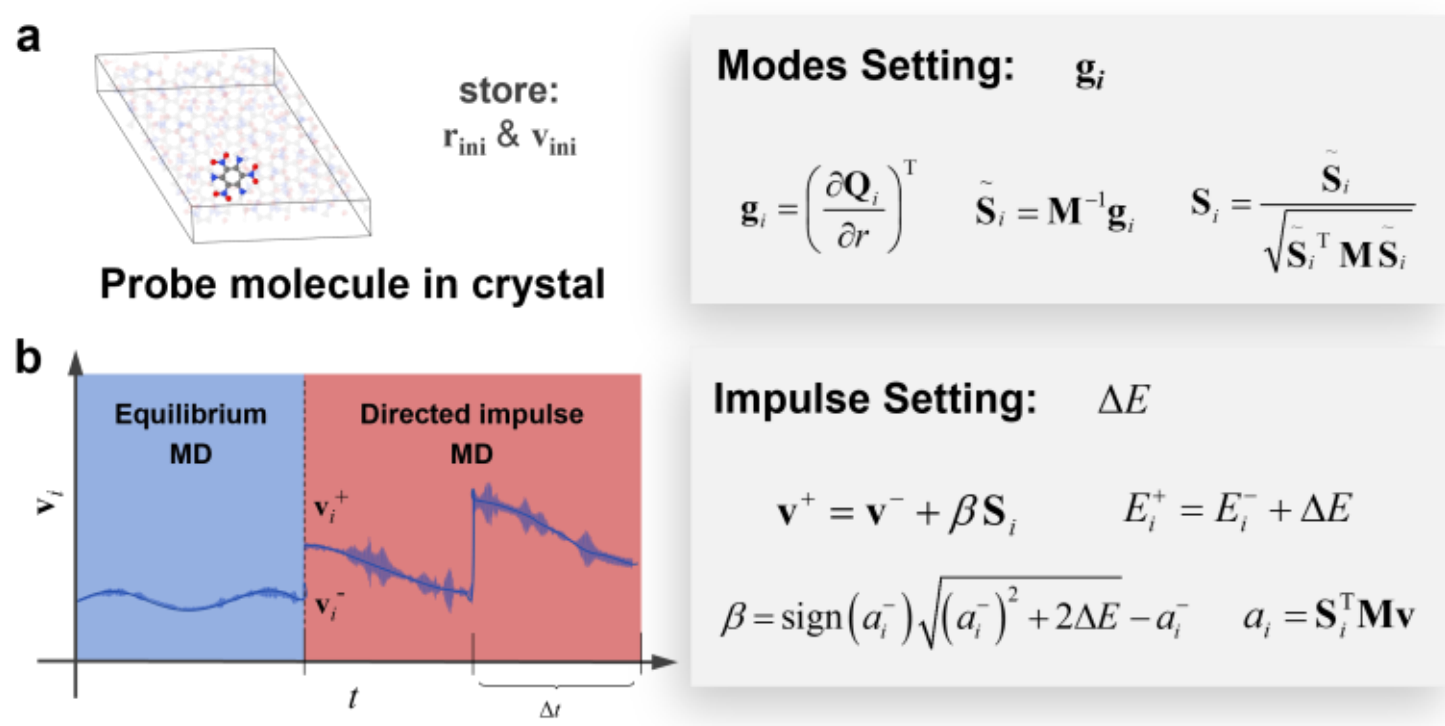

**Fig.** 2 **The Schematic diagram of Directed impulsive molecular dynamics.** (a) Definition of selected vibrational modes using physically interpretable internal coordinates. These modes can serve either as excited modes for impulsive energy injections or as monitored modes for tracking subsequent energy redistribution. (b). Energy impulse Setting. A fixed energy increment is prescribed for each impulse, and the velocities of the participating atoms are updated with mode-specific weights derived from the definition of corresponding internal coordinate mode, ensuring identical energy injection in every impulse. In contrast to equilibrium MD, DIMD selectively injects energy into a prescribed degree of freedom, while repeated impulses amplify the resulting nonequilibrium energy-routing response.

## Initial bond cleavage depends on the excited vibrational mode

Under equal-energy impulse excitation, the onset of bond cleavage was strongly mode-dependent, as evidenced by the markedly different numbers of impulse required to trigger the first bond cleavage event across the eleven excited vibrational modes (Fig. 3a). To quantify these differences, we calculated the survival probability, defined as the fraction of trajectories remaining intact, as a function of impulse number for each mode (Fig. 3b). We further characterized the relative propensity for bond cleavage using a random advantage index (RAI, the expression is an equivalent form of the Common Language Effect Size.[28]) in Fig. 3c. The RAI aggregates the pairwise ordering of the impulse counts between a given mode and all other modes, thereby quantifying its overall relative advantage in triggering an earlier bond-cleavage event. An RAI above 0 indicates a general tendency to induce cleavage earlier than the other modes, whereas an RAI below 0 indicates a comparatively delayed response.

Ring-breathing excitation produced the most rapid bond cleavage response, as evidenced by its early event distribution and rapid decline in survival probability. By contrast, $NO_2$ wagging excitation produced the strongest suppression of bond cleavage, with the latest event distribution and the largest fraction of trajectories remaining intact at the end of the observation window. $NO_2$ torsional and ring out-of-plane excitation also delayed cleavage substantially relative to most $NH_2$ angular modes. Because the energy deposited after a given number of impulses was identical across all excitation modes, this RAI ordering demonstrates that different excited vibrational modes deliver the injected energy to bond-stretching coordinates with different efficiencies.

The initially excited mode also redistributed the first bond cleavage events between the N–H and C–$NO_2$ channels (Fig. 3d). N–H cleavage is kinetically more accessible than C–$NO_2$ cleavage, as indicated by its lower activation energy. DIMD nevertheless revealed that excitation of certain modes can override this intrinsic thermal preference. Ring-breathing excitation not only triggered bond cleavage most efficiently but also showed a pronounced preference for the higher-barrier C–$NO_2$ channel. In contrast, ring out-of-plane excitation preferentially induced N–H cleavage and produced a substantially slower overall response than the other two ring-deformation modes. Thus, even structurally related excited vibrational modes can produce distinct bond cleavage events, suggesting that they establish different microscopic tendencies of energy flow.

These observations reveal two distinct aspects of mode dependence. First, the initially excited mode governs the onset of the first bond-cleavage event, as reflected by the number of impulses required. Second, it dictates the partitioning of the deposited energy between the N–H and C–$NO_2$ channels.

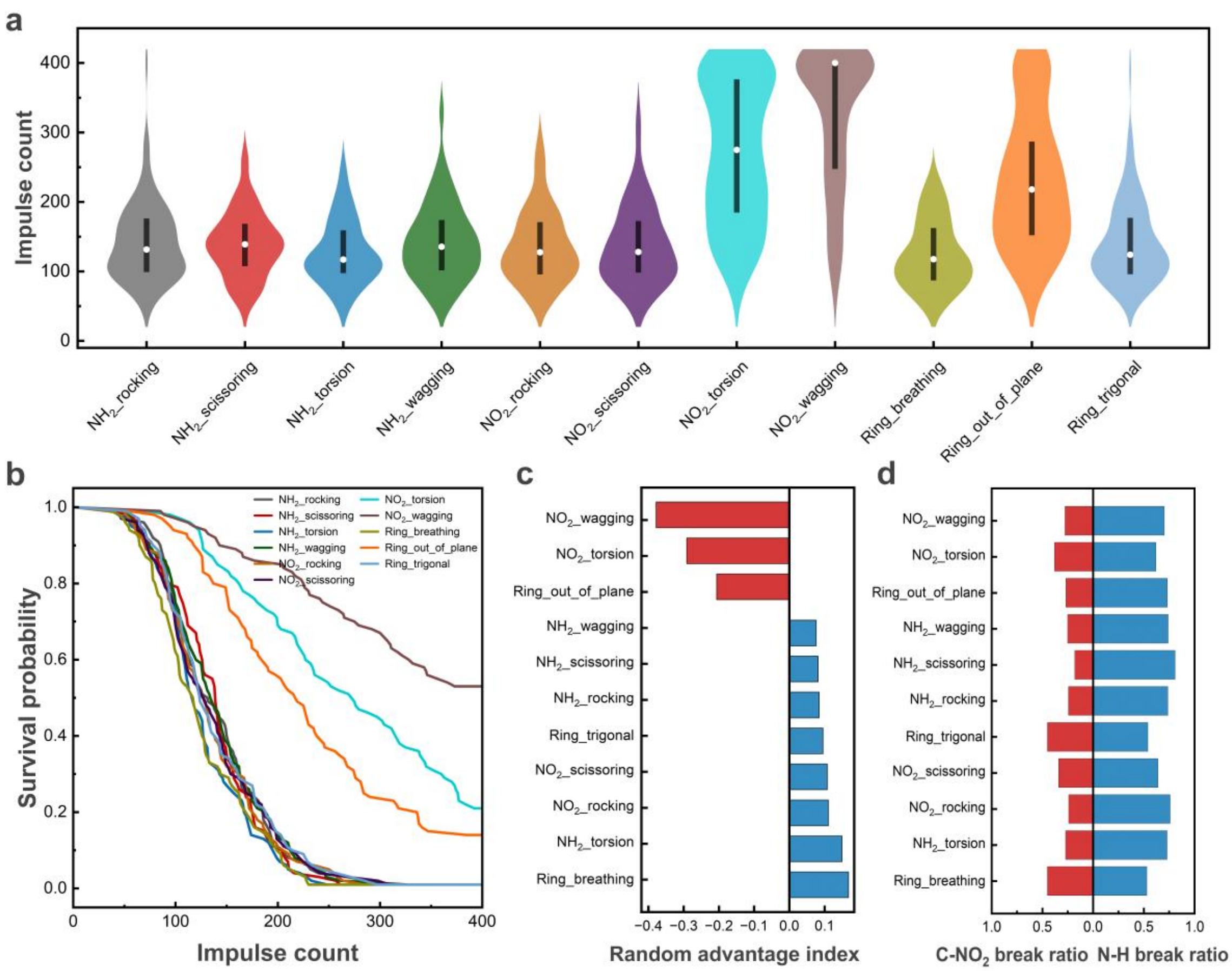

**Fig.** 3 **Quantitative analysis of the dependence of initial bond cleavage on the excited vibrational mode.** (a) Distributions of the number of equal-energy impulses required to trigger the first bond cleavage event for the eleven excited vibrational modes. The embedded box plots indicate the median and interquartile range. Trajectories that remained intact throughout the 400-segment observation window are shown at the maximum impulse count. (b) Mode-dependent survival probability as a function of impulse count. Survival probability is defined as the fraction of trajectories in which no bond cleavage event has occurred. (c) RAI summarizing the relative ordering of the first bond cleavage distributions across the excitation modes. (d) Fractions of the observed first bond cleavage events involving C–$NO_2$ cleavage (red) and N–H cleavage (blue) under each excited mode. Statistics for each mode were obtained from 100 independent initial configurations subjected to the same DIMD loading protocol.

## Mode-dependent early nonequilibrium energy flow

Motivated by the selective energy-transfer pathways proposed in multiphonon up-pumping theory, we examined the mode-resolved energy evolution before the onset of bond cleavage event. As defined in the mode-resolved energy projection section, instantaneous velocities and structural displacements were projected onto the steering vectors. The projected kinetic term $E_i(t)$ measures generalized motion along mode $i$, whereas the projected potential term $U_i(t)$ measures deformation along that direction using a fixed reference curvature from an embedded molecular Hessian.[29] Figs. S4–S14 show the time-dependent projected energies and their time-integrated changes under excitation of each mode. Although the instantaneous projected energies exhibit rapid oscillations, their integrated changes reveal persistent tendencies for energy accumulation or depletion. The response to the first few impulses is strongly affected by the initial vibrational phase, precluding robust comparisons among excited modes. Over longer timescales progressive thermalization, substantial structural deformation and bond cleavage modify the initial nonequilibrium energy-flow tendencies and alter the number of trajectories contributing to the ensemble average. We therefore selected 0.5–2 ps as an empirical early nonequilibrium window, within which the integrated energy changes evolve approximately linearly. By 0.5 ps, the influence of the initial oscillations of the projected energy had largely diminished, whereas no bond-cleavage event occurred in any trajectory before 2 ps. Within this interval, energy flow remained predominantly nonequilibrium, and the influence of the initially excited mode remained physically interpretable.

We first extracted the time-integrated changes in the self-projected kinetic and potential energies of each initially excited mode (Fig. S15). These quantities characterize whether, under nonequilibrium excitation, the energy deposited into a mode tends to remain as kinetic motion, be converted into deformation, or leave the initially excited mode. None of the modes considered here is itself defined as a direct N–H or C–$NO_2$ bond-stretching coordinate.

Under equal-energy excitation, the onset of bond cleavage is governed not only by the intrinsic accessibility of the reaction pathway, such as the bond strength and activation barrier, but also by the efficiency with which the deposited energy reaches and accumulates in the corresponding bond-stretching coordinate. Before cleavage occurs, the injected energy may remain localized within the initially excited mode or be redistributed through anharmonic coupling into deformation-associated degrees of freedom.[5-8,15] Characterizing their nonequilibrium self-responses therefore provides insight into distinguish deformation-associated energy accumulation and kinetic-energy retention. Because the instantaneous projected energies contain rapid vibrational oscillations, we quantified these persistent tendencies using the slopes of their time-integrated changes over the 0.5–2 ps window. For a projected-energy change $\Delta X_i(t)$, its time-integrated response is defined as

$$I(\Delta X_i; t) = \int_0^t \Delta X_i(t')\,\mathrm{d}t', \tag{1}$$

where $X_i$ denotes the potential or kinetic energies projected onto the vibrational mode $i$ and $k_I(\Delta X_i)$ denotes the slope obtained by fitting $I(\Delta X_i; t)$ over 0.5–2 ps with a linear function constrained to pass through the origin. This slope characterizes the persistent early-time tendency for energy accumulation under dynamics conditions.

On this basis, we introduced two complementary descriptors, termed the soft-like and trap indices. The first descriptor, termed the soft-like index, is defined as

$$\text{soft-like}_i = \frac{k_I(\Delta U_{i,\text{self}})}{k_I(\Delta E_{i,\text{self}})}, \tag{2}$$

The soft-like index measures whether energy retained by the excited mode is expressed preferentially as a deformation-associated potential term or as a vibrational kinetic term. A larger value indicates a stronger tendency towards structural deformation, whereas a smaller value indicates that the response remains predominantly kinetic. The second descriptor, termed the trap index, is defined for excited mode $i$ and bond-stretching coordinate $j$ as

$$\text{trap}_{i,j} = \frac{k_I(\Delta E_{i,\text{self}})}{k_I(\Delta E_{j|i})},\ j \in \{\text{N}-\text{H},\ \text{C}-\text{NO}_2\}, \tag{3}$$

The trap index therefore measures the extent to which kinetic energy remains associated with the initially excited mode relative to its delivery to a bond-stretching coordinate. A larger value indicates stronger energy retention and less efficient transfer to the corresponding bond-stretching coordinate, whereas a smaller value indicates more effective energy delivery to that coordinate. Among the modes examined, $NO_2$ torsion exhibits the strongest soft-like and energy-trapping characteristics (Fig. 4), indicating that excited energy transferred into $NO_2$ torsion is less likely to trigger bond cleavage.

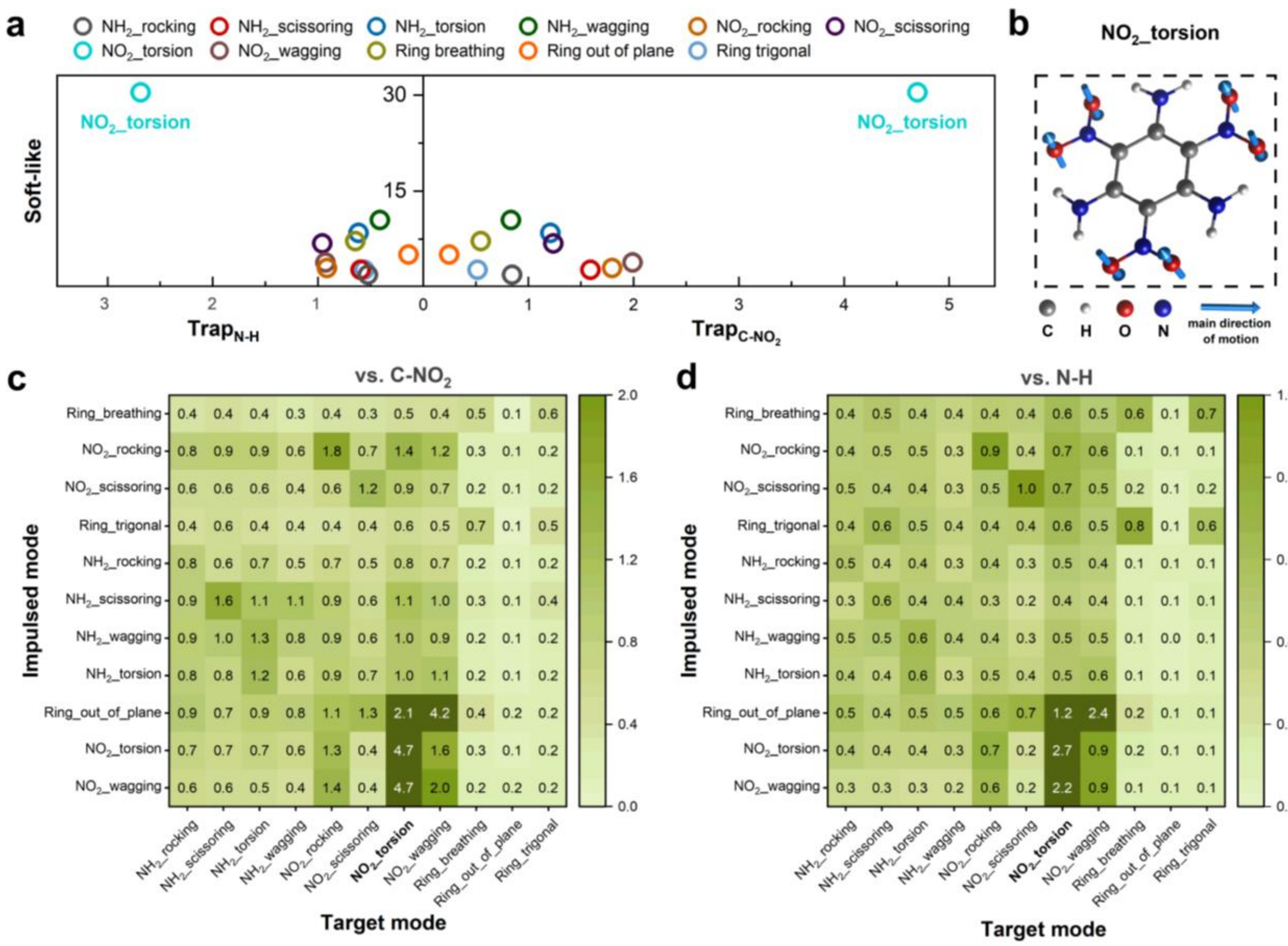

**Fig.** 4 **Early nonequilibrium energy routing.** (a) Two-index map of mode-dependent early nonequilibrium energy flow. The soft-like and trap indices were calculated from the slopes of the time-integrated projected-energy changes over the 0.5–2 ps early nonequilibrium window. The vertical axis shows the soft-like index, which characterizes the tendency of an excited mode to accommodate energy as structural deformation relative to vibrational kinetic motion. The horizontal axes show the trap indices with respect to the N–H and C–$NO_2$ bond-stretching coordinates. Each marker represents one of the eleven excited vibrational modes. (b) Schematic diagram of main motion for the $NO_2$

torsion mode. (c, d) Heatmaps show the relative energy-inflow indices from each initially excited mode (rows) into the target modes (columns), normalized by the energy delivered to the (c) C–$NO_2$ and (d) N–H stretching coordinates. Each value represents the ratio of the slope of the time-integrated kinetic-energy response of the target mode to that of the corresponding bond cleavage coordinate over the 0.5–2 ps window. The excited modes (rows) are ordered from top to bottom by increasing difficulty in triggering the bond cleavage event, *i.e.*, RAI.

To quantify energy transfer toward the reactive bond-stretching coordinates, we further compared the energy accumulation of each target mode in different excited modes with that accumulated in the corresponding N–H or C–$NO_2$ stretching coordinate. For an initially excited mode $i$, a target mode $m$, and a bond-breaking coordinate $j$, we define the relative energy-inflow index as

$$\mathrm{inflow}_{i\to m}^{(j)} \frac{k_I(\Delta E_{m|i})}{k_I(\Delta E_{j|i})},\ j \in \{\mathrm{N-H},\ \mathrm{C-NO_2}\}, \tag{4}$$

where $k_I(\Delta E_{m|i})$ is the slope of the time-integrated kinetic-energy response of target mode $m$ under excitation of mode $i$. A larger inflow index indicates that energy is preferentially accumulated in the target mode rather than other modes. Fig. 4c and d maps these relative energy-inflow index with respect to the C–$NO_2$ (Fig. 4c) and N–H (Fig. 4d) stretching coordinates. The excited modes are ordered from top to bottom by increasing difficulty in triggering bond cleavage. $NO_2$ torsion acts as a prominent energy-receiving mode under $NO_2$ wagging and ring out-of-plane modes, both of which also exhibit slow bond-cleavage responses. The particularly low reactivity of $NO_2$ wagging can be rationalized by its highly concentrated energy inflow into modes that are ineffective at promoting cleavage, especially $NO_2$ torsion and $NO_2$ wagging. By comparison, energy deposited into $NO_2$ torsion is distributed more broadly among several other modes, consistent with its less pronounced suppression of bond cleavage event relative to $NO_2$ wagging. Ring-breathing mode exhibits distinct behavior. Relative inflow into other modes is generally weak when normalized by energy delivery to the C–$NO_2$ stretching coordinate, but becomes more pronounced when normalized by delivery to the N–H stretching coordinate. This contrast indicates that ring-breathing mode preferentially channels energy towards C–$NO_2$ stretching rather than N–H stretching, consistent with its observed selectivity for C–$NO_2$ cleavage. Thus, sufficiently directional nonequilibrium energy routing can favor cleavage of a thermally less accessible bond, even when that bond has a higher apparent activation barrier.

## Discussion

In summary, by combining a vibrationally faithful neural-network interatomic potential with directed impulsive molecular dynamics, we tracked the nonequilibrium redistribution of externally deposited energy and its connection to initial bond cleavage. We demonstrate that, even with identical amounts of deposited energy, the vibrational form through which energy initially enters the system substantially affects the first bond cleavage event. The dependence of bond cleavage on the excited mode originates from early nonequilibrium energy-transfer pathways established before the reaction occurs. Energy deposited into a vibrational degree of freedom does not immediately disperse randomly but instead follows mode-specific routes. These complex routes can be understood as a competition between two transfer processes. On the one hand, energy is directed towards a specific bond-stretching coordinate, thereby shortening the time to first cleavage and increasing the fraction of the corresponding bond cleavage channel. On the other hand, energy is captured by soft-like or energy-trapping modes. Some modes accommodate substantial energy through structural deformation or retain energy that cannot be efficiently transferred to the bond-stretching coordinate. These modes therefore reduce the efficiency with which reactive degrees of freedom are energized and consequently delay bond cleavage. In TATB, ring-breathing excitation preferentially channels energy toward C–$NO_2$ stretching while bypassing the $NO_2$ torsional mode, which exhibits high soft-like and trap indices. This energy-routing pathway promotes earlier bond cleavage and favors C–$NO_2$ rupture despite its higher apparent activation energy relative to N–H cleavage. These findings connect vibrational motion with the energy-routing pathways that govern the onset and selectivity of the initial chemical response. It provides a microscopic basis for understanding the early stages of initiation and identifies the redistribution of energy among vibrational modes as an unnegligible mechanism through which molecular structure can influence initiation sensitivity.

## Methods

### NNAP construction of TATB

The NNAP database was assembled progressively through molecule, crystal, pyrolysis and shock stages, following the knowledge-guided workflow developed in our previous work.[18] The crystal stage treated TATB-specific structural fidelity as the iterative convergence criterion. The sampling protocols specified in Fig. s1 were adopted for all other stages, and no explicit iterative convergence criterion were defined.

To evaluate TATB-specific structural fidelity, DFT and NNAP structures were optimized under hydrostatic pressures from 0 to 50 GPa. For each optimized structure, aromatic-carbon centroids were used to define molecular planes and interlayer spacing along the layer normal. Intralayer hydrogen bonds were characterized by H⋯O distances and N-H⋯O angles, and the nearest interlayer H⋯O contacts were collected separately. Hydrogen-bond contacts were counted using H⋯O < 2.8 Å and N-H⋯O > 110°. Means and standard deviations of all descriptors were compared between DFT and the current NNAP, and their pressure-integrated mean absolute errors were used to determine whether additional crystal configurations remained necessary.

During the NNAP iteration process, we trained a 10-member committee of NNAP models using 10-fold cross-validation. The filtering protocol for sampled configurations, dataset partitioning strategy, and functional form of the basis set were consistent with those adopted in our previous work.[18] Variations in the basis set parameters of this work were attributed to $r_{cut}^{(2b)}$ = 6.5 Å and bond parameters were selected as $b_0$ = 0.9.

### DFT calculations and structural optimization

All energies and atomic forces used for training were obtained using atomic simulation environment (ASE)[30] and Vienna ab-initio simulation package (VASP)[31]. Before DFT calculations, initial magnetic moments were assigned to each configuration using crystal hamiltonian graph neural network (CHGNet)[32] to facilitate charge and spin convergence. The Perdew–Burke–Ernzerhof (PBE)[33] exchange–correlation functional was employed, combined with Tkatchenko–Scheffler (TS)[34] dispersion correction to account for van der Waals interactions.

For TATB crystal calculations, convergence tests determined that an energy cutoff of 700 eV was sufficient for the plane-wave basis set. The *k*-point spacing was set to 0.25 $Å^{-1}$ for bulk, while for isolated systems placed in a 10 Å vacuum layer, a *k*-point spacing of 50 $Å^{-1}$ was used.

All structural optimizations were performed using ASE interfaced with VASP, large-scale atomic/molecular massively parallel simulator (LAMMPS)[35]. DFT calculations were conducted at the PBE+D3(BJ)[36] level of theory. Geometry relaxations were carried out using the CG algorithm[37] in VASP and LAMMPS, combined with UnitCellFilter[38] in ASE to relax cell vectors. Convergence was achieved when the maximum atomic force was below 0.01 eV·$Å^{-1}$.

### Infrared spectra calculation

Infrared spectra of crystalline TATB were calculated using NNAP, ReaxFF[23] and EMFF.[24] A periodic 3×3×3 TATB supercell was first minimized and then equilibrated at 300 K for 5 ps under NPT ensemble at zero pressure, followed by 5 ps under NVT ensemble. Production trajectories were subsequently generated under 100 ps NVE ensemble using a timestep of 0.5 fs.

For the ReaxFF simulations, both the interatomic forces and instantaneous atomic charges were evaluated within the ReaxFF framework. For the NNAP and EMFF simulations, the corresponding potentials were used to describe the atomic interactions and dynamical evolution, while instantaneous atomic charges were determined at every timestep using the charge-equilibration (QEq)[39] method with the parameter set from ReaxFF.[23]

The instantaneous total dipole moment was calculated as

$$\mathbf{M}(t) = \sum_{i=1}^{N} q_i\,(t)\mathbf{r}_i(t), \tag{5}$$

where $q_i(t)$ and $\mathbf{r}_i(t)$ are the instantaneous partial charge and unwrapped position of atom $i$, respectively. The dipole velocity was evaluated using the central finite difference,

$$\dot{\mathbf{M}}(t_n) = \frac{\mathbf{M}(t_{n+1}) - \mathbf{M}(t_{n-1})}{2\Delta t}. \tag{6}$$

The dipole-velocity trajectories were divided into 10 ps segments.[40,41]

$$\tilde{\dot{\mathbf{M}}}_k(\omega) = \int_0^{T_\mathrm{w}} \dot{\mathbf{M}}_k(t) e^{-i\omega t}\,\mathrm{d}t, \tag{7}$$

and the classical infrared spectrum was obtained from

$$I_\mathrm{cl}(\omega) \propto \left\langle \left| \tilde{\dot{\mathbf{M}}}_k(\omega) \right|^2 \right\rangle_k. \tag{8}$$

A harmonic quantum-correction factor[42] was applied to approximate nuclear quantum effects with $T = 300$ K,

$$I(\omega) = \frac{\beta\hbar\omega}{1-\exp(-\beta\hbar\omega)} I_\mathrm{cl}(\omega),\ \beta = \frac{1}{k_\mathrm{B}T}. \tag{9}$$

**Pyrolysis simulations and activation energy calculations**

Pyrolysis simulations were performed using NNAP on a supercell of TATB (456 molecules) under the NVT ensemble at 1,400–1,600 K in increments of 50 K, each for 5 ns with a timestep of 0.1 fs. The initial intramolecular N–H and C–$NO_2$ bond networks were constructed from predefined atom-index pairs. For each initially bonded atom pair $i$ and $j$, the instantaneous interatomic distance was calculated using the minimum-image convention. A bond-specific distance cutoff was defined as

$$r_{ij}^\mathrm{cut} = 1.3\left(r_i^\mathrm{cov} + r_j^\mathrm{cov}\right), \tag{10}$$

where $r_i^\mathrm{cov}$ and $r_j^\mathrm{cov}$ are the covalent radii of the corresponding elements, the factor of 1.3 was adopted from a previously reported[43] bond-identification criterion for high-temperature reactive MD simulations of energetic materials. A bond was classified as intact when $r_{ij} \le r_{ij}^\mathrm{cut}$ and as cleaved when $r_{ij} > r_{ij}^\mathrm{cut}$. The surviving fraction of each bond type $\alpha$, where $\alpha$ denotes N–H or C–$NO_2$, was calculated as

$$S_\alpha(t) = \frac{N_\alpha^\mathrm{intact}(t)}{N_\alpha^\mathrm{intact}(0)}. \tag{11}$$

The temperature-dependent cleavage rate constant $k_\alpha(T)$ was obtained by fitting the decay of the surviving fraction to a first-order exponential model,

$$S_\alpha(t) = \exp[-k_\alpha(T)t]. \tag{12}$$

The apparent activation energy $E_{\mathrm{a},\alpha}$ for each bond-cleavage channel was subsequently obtained from the Arrhenius relation,

$$k_\alpha(T) = A_\alpha \exp\left(-\frac{E_{\mathrm{a},\alpha}}{RT}\right), \tag{13}$$

where $A_\alpha$ is the pre-exponential factor and $R$ is the gas constant.

**DIMD and vibrational mode setting**

The DIMD simulation cell contained 1,296 atoms (54 TATB molecules) under periodic boundary conditions. A single 24-atom probe molecule was used to define the internal-coordinate steering directions and the bond network monitored during the simulations. 100 initial configurations were sampled from an equilibrated trajectory at 800 K. Dynamics between impulses used the NNAP in LAMMPS. The impulse interval was set to 50 fs according to the initial decay timescale of the total velocity autocorrelation function.

The DIMD implementation supports four types of internal coordinates,

$$q_{ij}^{\mathrm{bond}} = \left|\mathbf{r}_j - \mathbf{r}_i\right|, \tag{14}$$

$$q_{ijk}^{\mathrm{angle}} = \cos\theta_{ijk} = \frac{(\mathbf{r}_i-\mathbf{r}_j)\cdot(\mathbf{r}_k-\mathbf{r}_j)}{|\mathbf{r}_i-\mathbf{r}_j||\mathbf{r}_k-\mathbf{r}_j|}, \tag{15}$$

$$q_{ijkl}^{\mathrm{dihedral}} = \phi_{ijkl} = \mathrm{atan2}\left[\hat{\mathbf{b}}_2\cdot(\mathbf{n}_1\times\mathbf{n}_2), \mathbf{n}_1\cdot\mathbf{n}_2\right], \tag{16}$$

where $\mathbf{b}_1 = \mathbf{r}_j - \mathbf{r}_i$, $\mathbf{b}_2 = \mathbf{r}_k - \mathbf{r}_j$, $\mathbf{b}_3 = \mathbf{r}_l - \mathbf{r}_k$, $\mathbf{n}_1 = \mathbf{b}_1 \times \mathbf{b}_2$, and $\mathbf{n}_2 = \mathbf{b}_2 \times \mathbf{b}_3$. Improper coordinates were evaluated using the same signed-dihedral definition, with atom $j$ specified as the central atom; for the index order $i, j, k, l$, the evaluated coordinate was $\phi_{kjil}$. All coordinate vectors were constructed using minimum-image atomic displacements.

A vibrational mode $m$ was expressed as a weighted linear combination of physically interpretable internal coordinate,

$$\mathbf{Q}_m(\mathbf{r}) = \sum_{p\in m} w_p\, q_p(\mathbf{r}). \tag{17}$$

Eleven vibrational modes were considered: $NH_2$ scissoring, wagging, rocking and torsion; $NO_2$ scissoring, wagging, rocking and torsion; ring breathing, ring trigonal and ring out-of-plane deformation. The scissoring modes were defined by the H–N–H or O–N–O angle coordinate. Rocking was represented by the antisymmetric combination of the two C–N–H or C–N–O angle coordinates, whereas wagging was described using an H–N–H–C or O–N–O–C improper coordinate. Torsion was constructed from the sum of two proper dihedral coordinates around the corresponding exocyclic C–N bond. Ring breathing was defined as the in-phase sum of the six ring C–C bond coordinates, ring trigonal deformation as an alternating-sign combination of the six ring-angle coordinates, and ring out-of-plane deformation as the sum of improper coordinates at three alternating ring sites. For each $NH_2$ or $NO_2$ mode, the three symmetry-related substituent sites constituted separate member modes, among which the assigned group impulse energy was distributed equally.

**Bond cleavage event criterion**

The DIMD was terminated when either the bond cleavage event criterion was met in the probe molecule or the preset maximum duration of 20 ps was reached. Bond cleavage was evaluated for the predefined bond network of the probe molecule using the same bond-specific distance cutoff described in Pyrolysis simulations and activation energy calculations section. Moreover, the cleavage condition was required to persist for at least 15 fs.

**Mode-resolved energy projection**

At each analyzed configuration, the mass-normalized steering vector $\mathbf{S}_i(t)$ of each internal-coordinate mode was reconstructed from the instantaneous molecular geometry. And the generalized displacement was calculated as

$$\xi_i(t) = \mathbf{S}_i(t)^T \mathbf{M}[\mathbf{r}(t) - \mathbf{r}_0]. \tag{18}$$

Because the steering vector varies with molecular geometry, the generalized velocity included both the instantaneous velocity projection and the time dependence of the steering vector:

$$\dot{\xi}_i(t) = \mathbf{S}_i(t)^{\mathrm{T}} \mathbf{M} \mathbf{v}(t) + \dot{\mathbf{S}}_i(t)^{\mathrm{T}} \mathbf{M}[\mathbf{r}(t) - \mathbf{r}_0]. \tag{19}$$

Here, $\dot{\mathbf{S}}_i(t)$ was estimated by backward finite difference between consecutive trajectory frames. The mode-resolved kinetic energy was then calculated as

$$E_i(t) = \frac{1}{2}\dot{\xi}_i(t)^2. \tag{20}$$

The mode-resolved potential energy was estimated within a local harmonic approximation. A Cartesian Hessian $\mathbf{H}_0$ was calculated for the probe molecule while retaining its interactions with the surrounding environment. The reference stiffness and projected potential energy of mode *i* were calculated as

$$\kappa_i^0 = \mathbf{S}_i(0)^{\mathrm{T}} \mathbf{H}_0 \mathbf{S}_i(0), \; U_i(t) = \frac{1}{2}\kappa_i^0 \xi_i(t)^{2.} \tag{21}$$

For group mode containing three symmetry-related member modes, the kinetic and potential energies were calculated separately for each member and then summed.

**Competing interests**

There are no conflicts to declare.

**Data availability**

The corresponding datasets will be made available upon acceptance of this work. Additional data that supports the findings of this study are available from the corresponding author upon reasonable request.

**Code availability**

The corresponding code will be made available upon acceptance of this work.

**Acknowledgements**

This work was supported by the Advanced Materials-National Science and Technology Major Project (Grant No. 2024ZD0606900), the Science Challenge Project. No.TZ2025013, the National Natural Science Foundation of China (Grants No. T2325004), and the Talent Hub for "AI+ New Materials" Basic Research.

**Author contributions**

L.W. and P.F.G. conceptualized and designed the research. G.C.D., X.N.H., R.S., L.W. and P.F.G. conducted the preliminary investigation. G.C.D., X.N.H., L.W. and P.F.G. analyzed the data and interpreted the results. G.C.D., Y.X.C., X.N.H., L.W., and P.F.G. jointly wrote the manuscript. All coauthor contributed to reading and editing the manuscript and approved the manuscript.

# Supplementary Information

## Directed impulsive molecular dynamics reveal energy-routing pathways of bond cleavage in TATB

*Guanchen Dong*[1,2,4,†], *Xiaona Huang*[5,†], *Yuxuan Chen*[1,3], *Linghua Tan*[2], *Guangcheng Yang*[2,4], *Rui Su*[6], *Lin Wang*[1,*], *Pengfei Guan*[1,*]

7. Advanced Interdisciplinary Science Research (AiR) Center, Ningbo Institute of Materials Technology and Engineering, Chinese Academy of Sciences, Ningbo 315201, China.
8. Nanjing University of Science and Technology, Nanjing, 210094, China.
9. Advanced Structural Ceramics Innovation Center, YongJiang Laboratory, Ningbo 315202, China.
10. China Academy of Engineering Physics·Institute of Chemical Materials, Mianyang, 621900, China.
11. School of Physics and Technology, Nanjing Normal University, Nanjing 210023, China.
12. Institute of Advanced Magnetic Materials, College of Materials & Environmental Engineering, Hangzhou Dianzi University, Hangzhou 310018, China.

[†]These authors contributed equally to this work.

**Corresponding Author**

* Lin Wang (wanglin@nimte.ac.cn); Pengfei Guan (pguan@nimte.ac.cn).

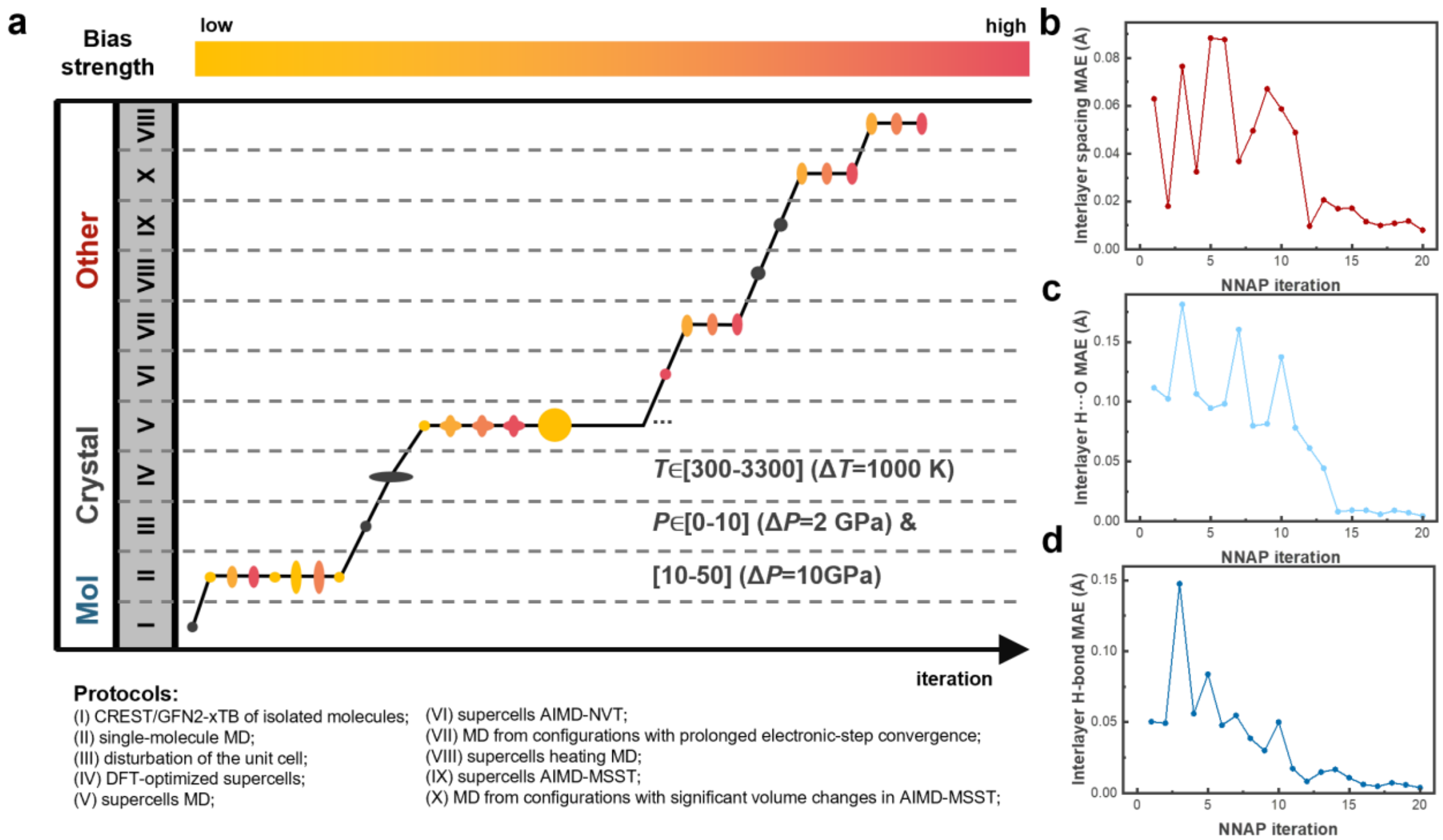

**Fig. S1** (a) NNAP sampling protocols and stage-specific iterative sampling workflow. Iterative convergence was evaluated from the mean absolute errors between NNAP- and DFT-optimized structures over 0–50 GPa in (b) interlayer spacing, (c) interlayer H···O distance and (d) intralayer hydrogen-bond distance.

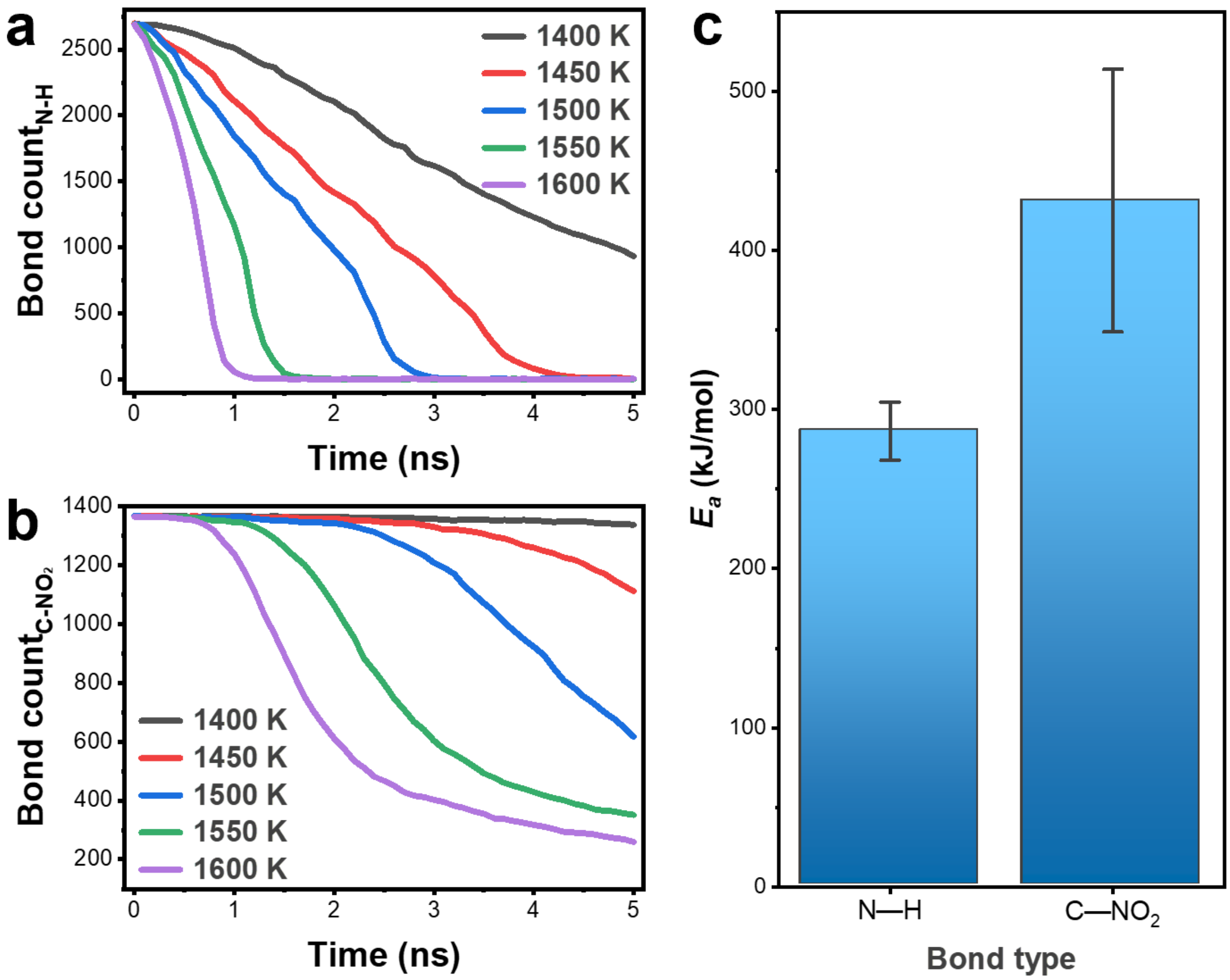

**Fig. S2** Temperature-dependent evolution of the number of intact (a) N–H and (b) C–$NO_2$ bonds, and (c) the corresponding apparent activation energies.

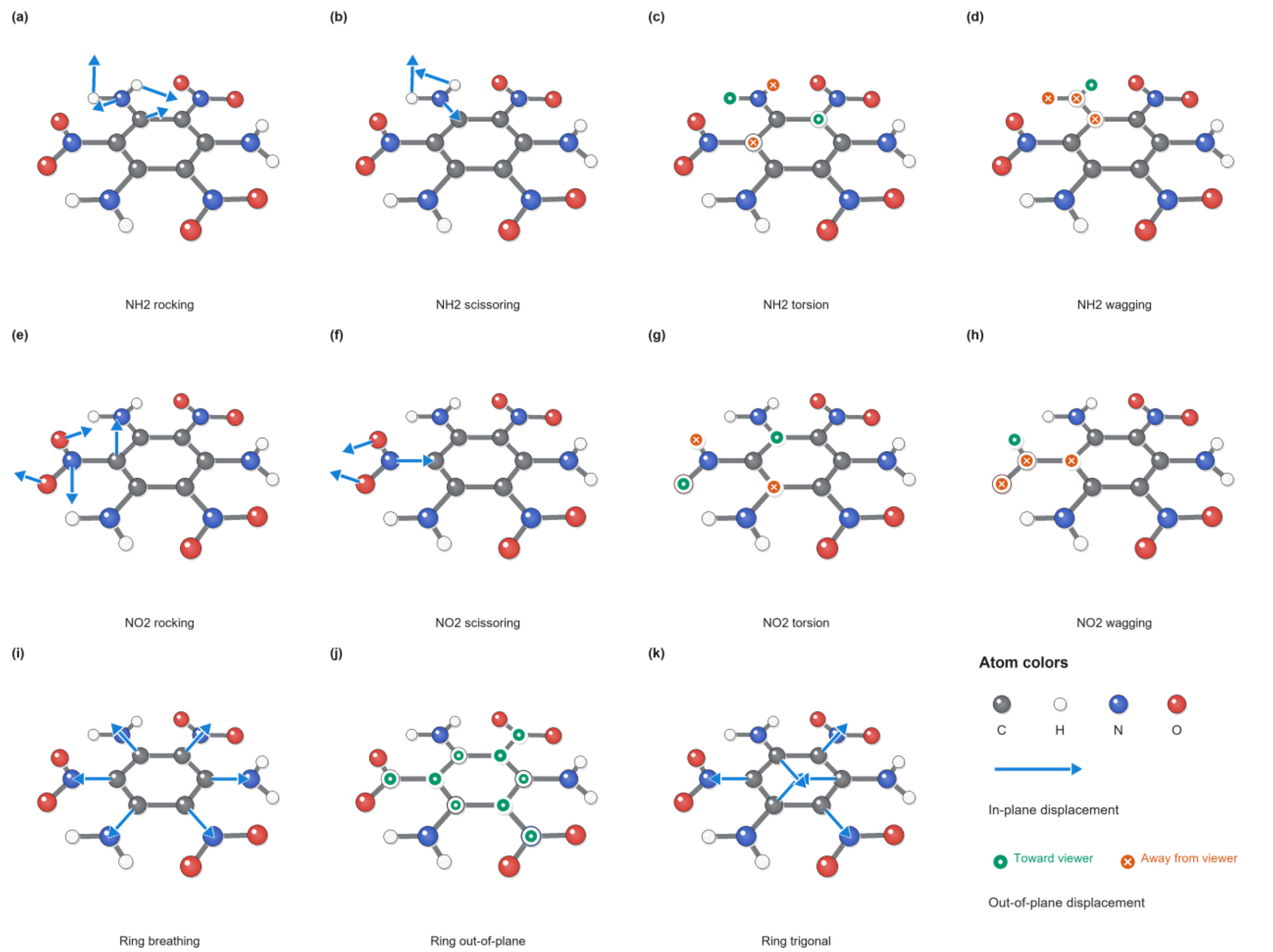

**Fig. S3** Internal-coordinate vibrational modes employed in DIMD simulations. Schematics of (a–d) $NH_2$ rocking, scissoring, torsion and wagging; (e–h) $NO_2$ rocking, scissoring, torsion and wagging; and (i–k) ring breathing, out-of-plane and trigonal modes. Blue arrows indicate in-plane atomic displacements, whereas green dots and orange crosses denote out-of-plane displacements toward and away from the viewer, respectively.

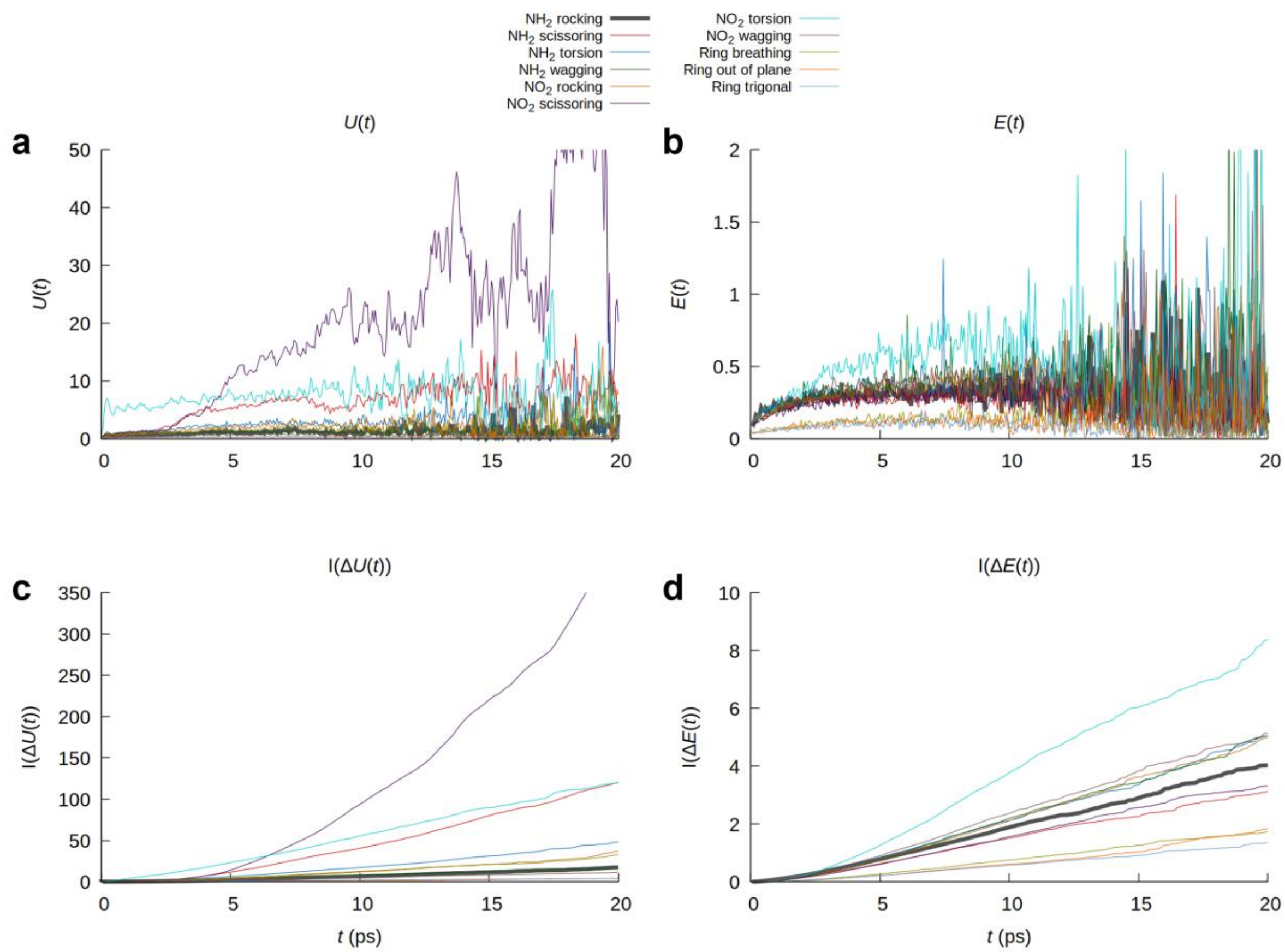

**Fig. S4** Mode-resolved projected energy under $NH_2$ rocking excitation. (a,b) Projected potential and kinetic energies; (c,d) their time-integrated changes.

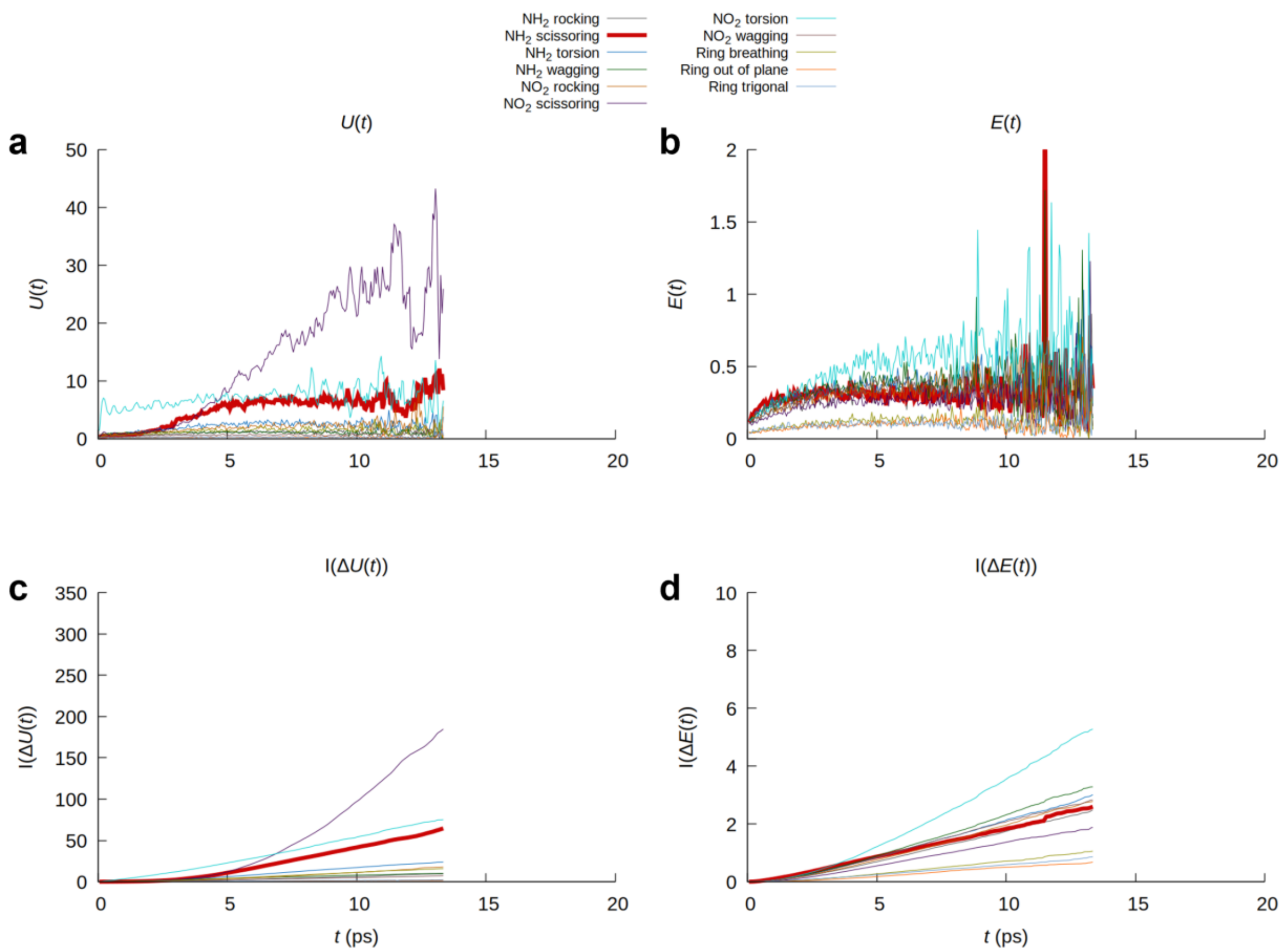

**Fig. S5** Mode-resolved projected energy under $NH_2$ scissoring excitation. (a,b) Projected potential and kinetic energies; (c,d) their time-integrated changes.

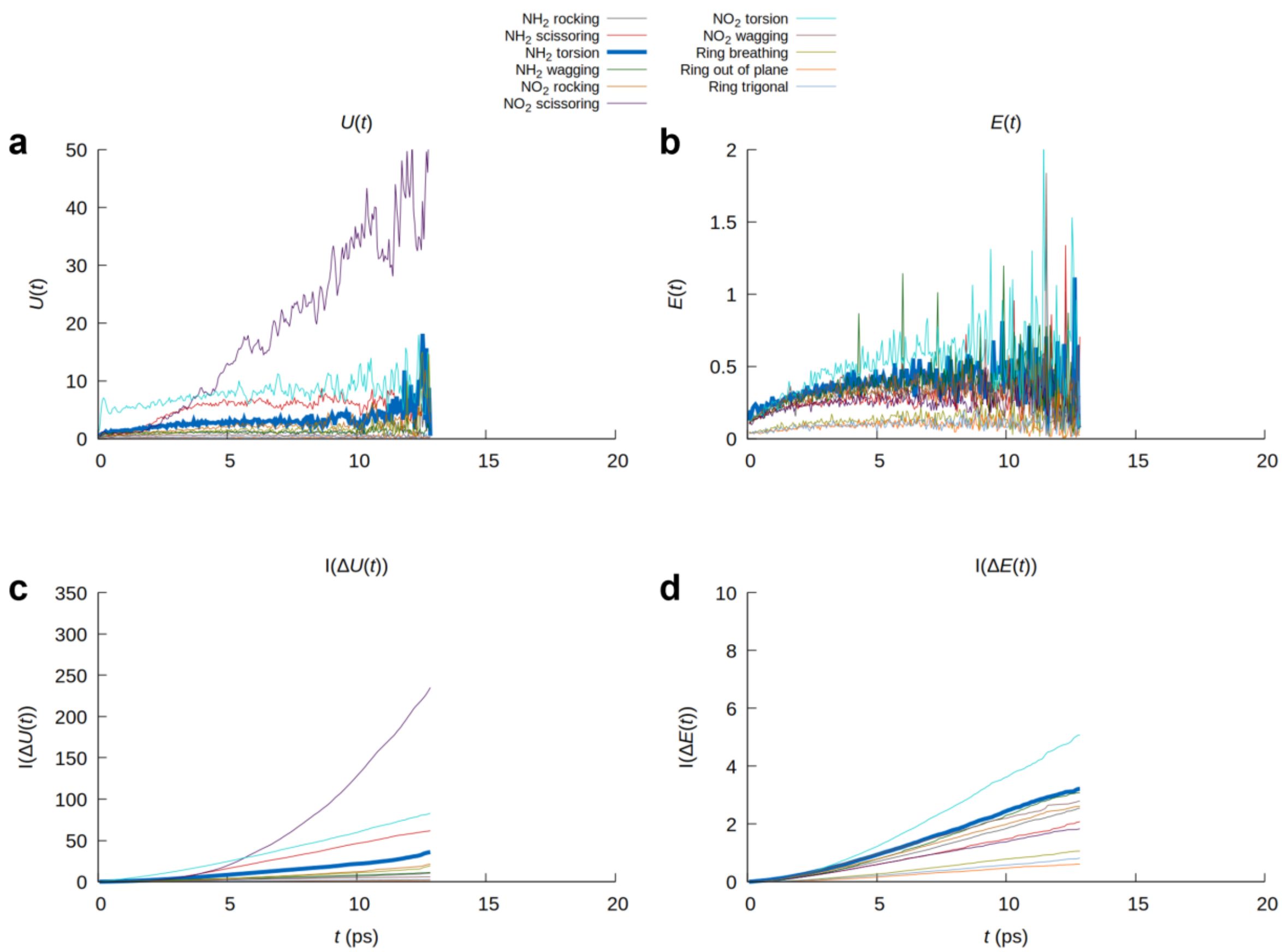

**Fig. S6** Mode-resolved projected energy under $NH_2$ torsion excitation. (a,b) Projected potential and kinetic energies; (c,d) their time-integrated changes.

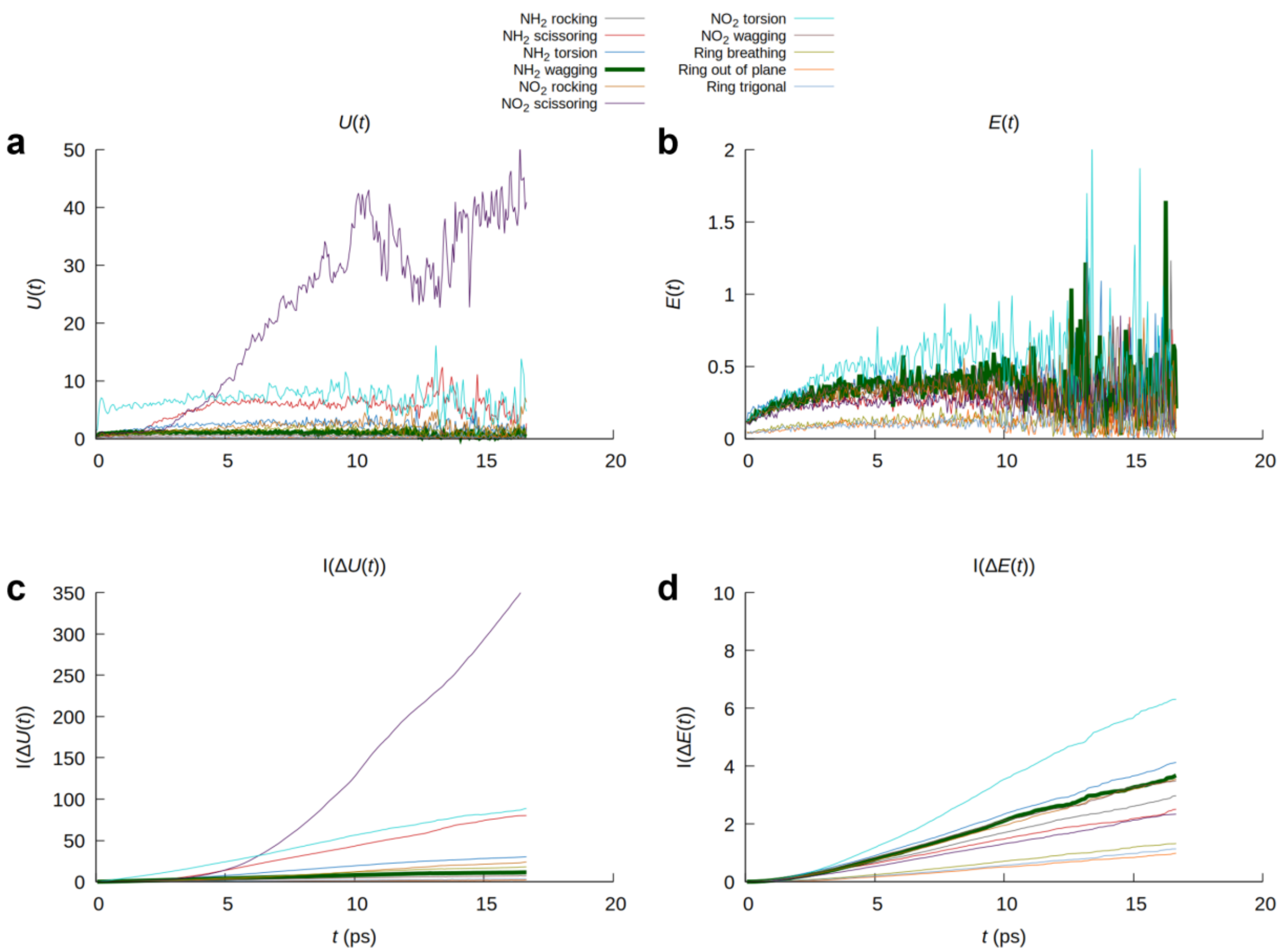

**Fig. S7** Mode-resolved projected energy under $NH_2$ wagging excitation. (a,b) Projected potential and kinetic energies; (c,d) their time-integrated changes.

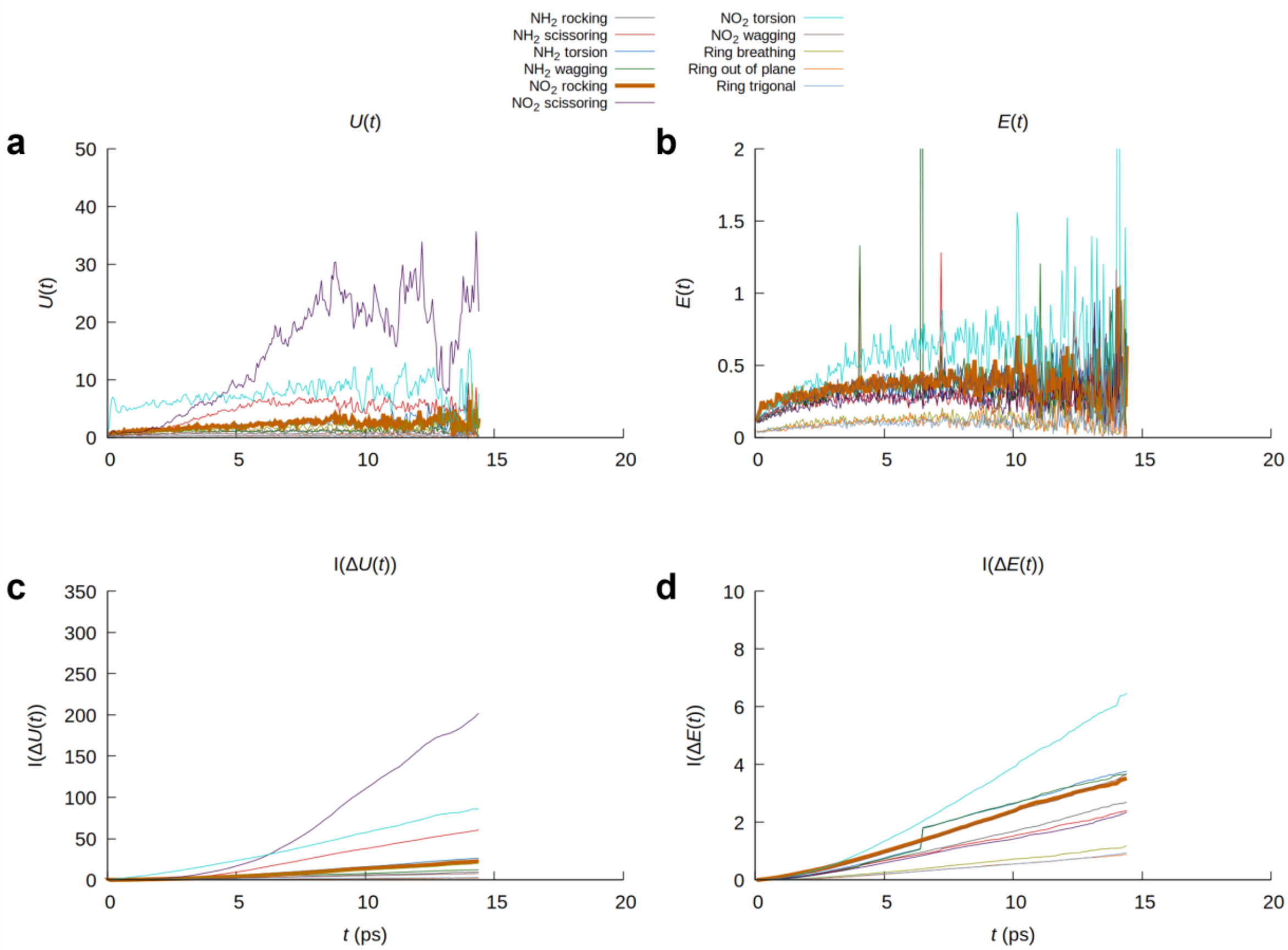

**Fig. S8** Mode-resolved projected energy under $NO_2$ rocking excitation. (a,b) Projected potential and kinetic energies; (c,d) their time-integrated changes.

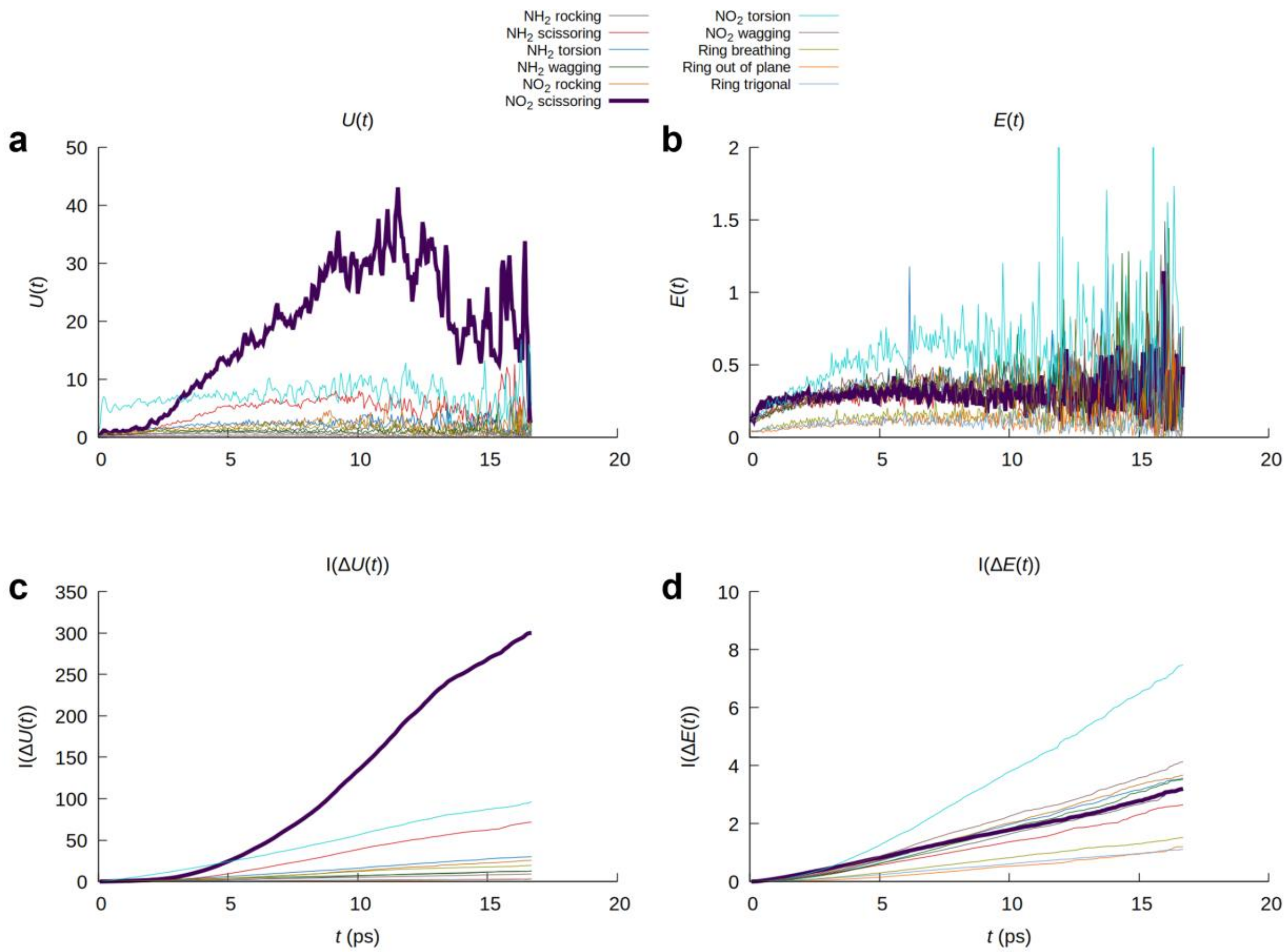

**Fig. S9** Mode-resolved projected energy under $NO_2$ scissoring excitation. (a,b) Projected potential and kinetic energies; (c,d) their time-integrated changes.

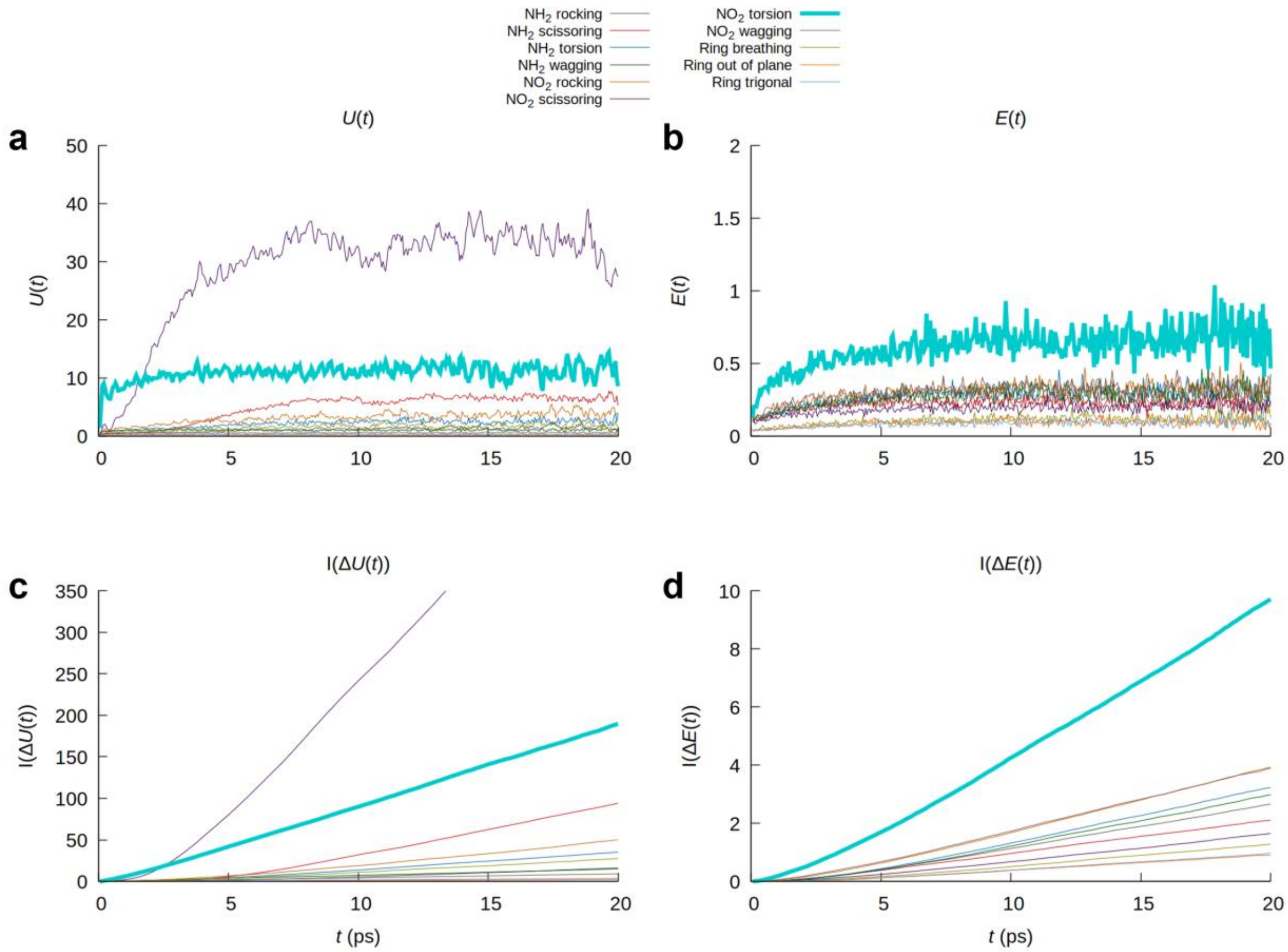

**Fig. S10** Mode-resolved projected energy under $NO_2$ torsion excitation. (a,b) Projected potential and kinetic energies; (c,d) their time-integrated changes.

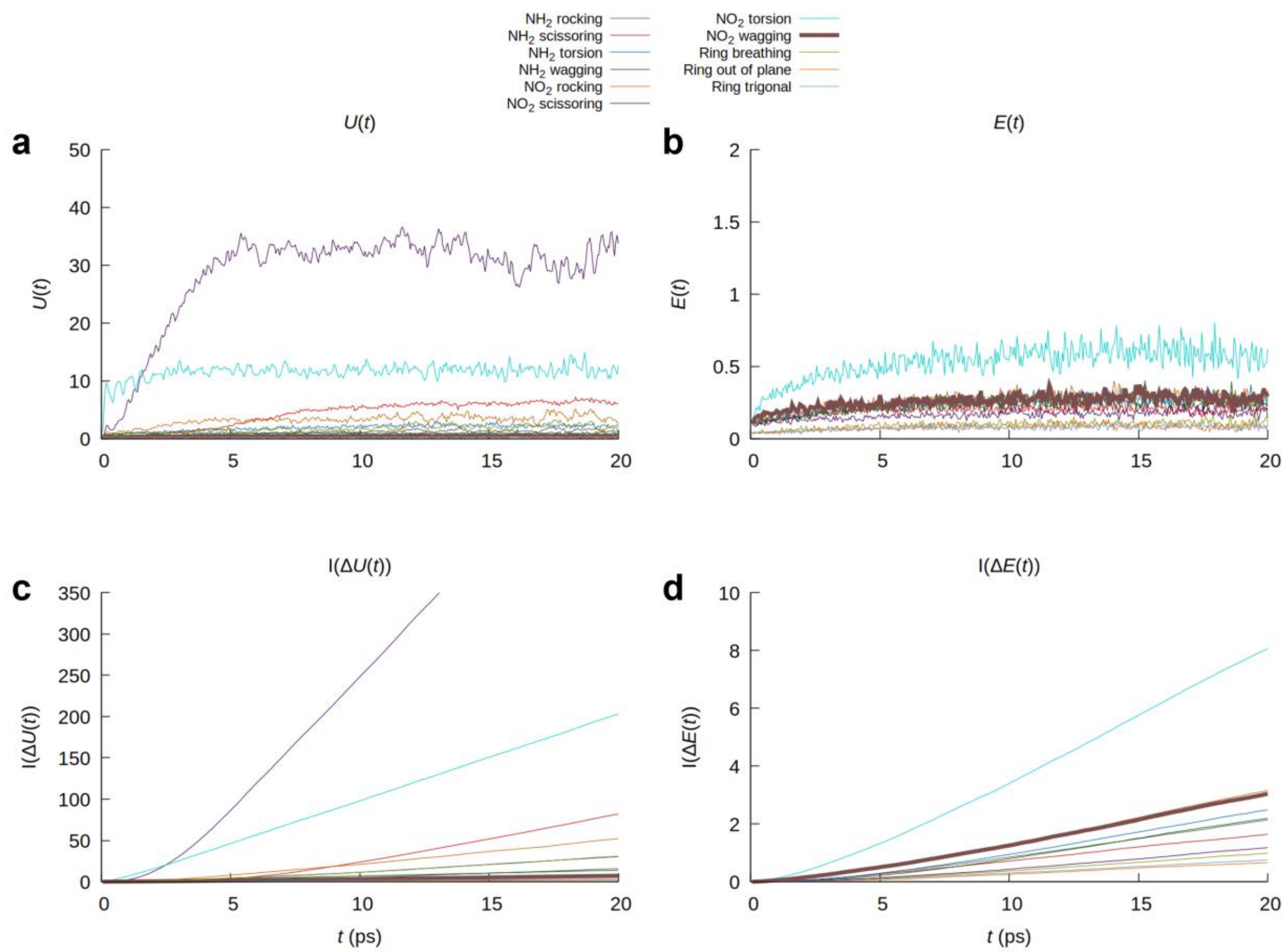

**Fig. S11** Mode-resolved projected energy under $NO_2$ wagging excitation. (a,b) Projected potential and kinetic energies; (c,d) their time-integrated changes.

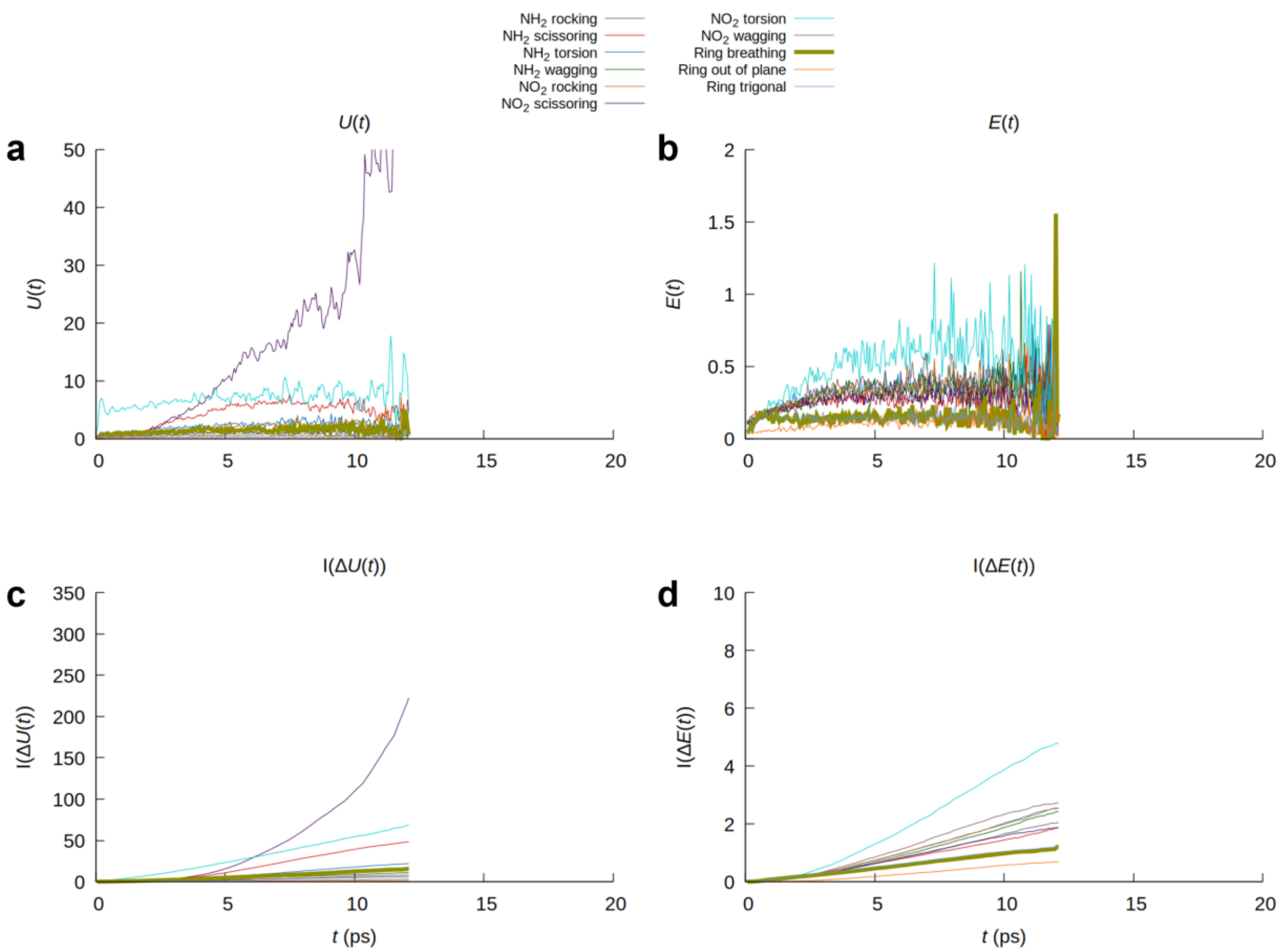

**Fig. S12** Mode-resolved projected energy under ring-breathing excitation. (a,b) Projected potential and kinetic energies; (c,d) their time-integrated changes.

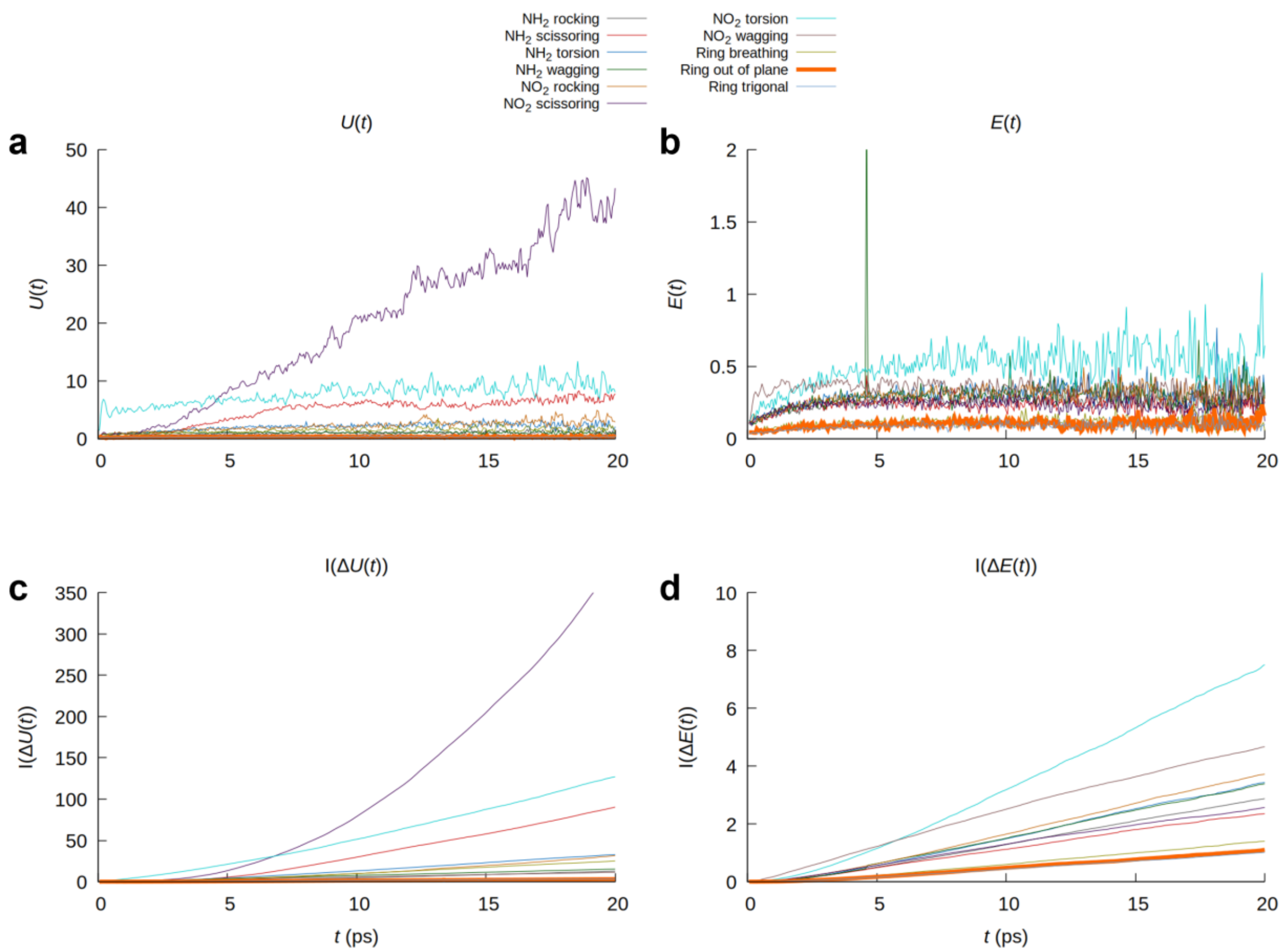

**Fig. S13** Mode-resolved projected energy under ring out-of-plane excitation. (a,b) Projected potential and kinetic energies; (c,d) their time-integrated changes.

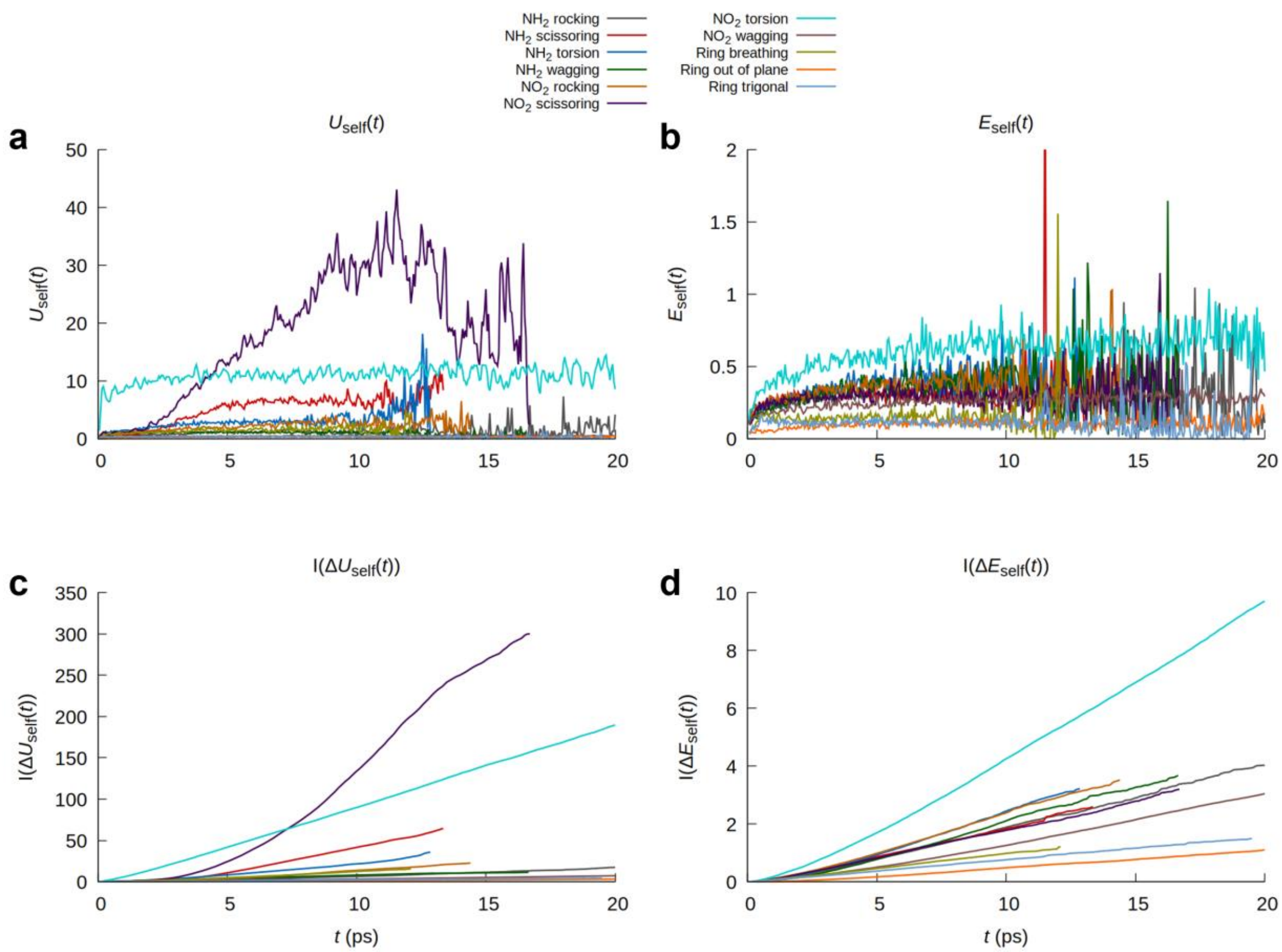

**Fig. S14** Self-projected energy responses of the eleven excited vibrational modes. (a,b) Self-projected potential and kinetic energies; (c,d) their time-integrated changes.